\documentclass[11pt, 3p]{elsarticle}
\makeatletter
\def\ps@pprintTitle{%
 \let\@oddhead\@empty
 \let\@evenhead\@empty
 \def\@oddfoot{\centerline{\thepage}}%
 \let\@evenfoot\@oddfoot}
\makeatother

\date{}
\def\texpsfig#1#2#3{\vbox{\kern #3\hbox{\includegraphics{#1}\kern #2}}\typeout{(#1)}}

\usepackage{url}
\usepackage[hidelinks]{hyperref}
\usepackage{bbm}
\usepackage{eurosym}                           
\usepackage[latin1]{inputenc}                  
\usepackage{url}                               
\usepackage{longtable}                         
\usepackage{array}                             

\usepackage{amsthm}
\usepackage{amsbsy}

\usepackage{placeins}  

\usepackage{amssymb}
\usepackage{amsmath}
\usepackage{bm}  

\usepackage{enumerate}

\usepackage{graphicx,color}
\usepackage{epsfig}
\usepackage[ruled,vlined]{algorithm2e}
\usepackage{multirow,bigdelim}
\usepackage[table]{xcolor}

\usepackage{natbib}
\setcitestyle{authoryear,open={(},close={)}}

\usepackage{cleveref}

\usepackage{diagbox} 

\usepackage{mathtools,eqparbox}

\usepackage{tikz}

\theoremstyle{plain}
\newtheorem{thm}{Theorem}[section]

\newtheorem{dfn}[thm]{Definition}

\newtheorem{rem}{Remark}[section]

\theoremstyle{remark}

\theoremstyle{plain}

\theoremstyle{definition}

\newcommand{\e}{{\rm e}}        
\def\R{\mathbb{ R}}             
\def\E{\mathbb{ E}}             
\def\Q{\mathbb{ Q}}             

\def\CVA{\mathrm{CVA}}

\def\F{\mathcal{F}}             

\renewcommand{\d}{{\rm d}}      

\def\e{{\mathrm{e}}}
\def\1{{\mathbbm{1}}}            

\def\EE{\mathrm{EE}}

\theoremstyle{plain}

\usepackage[margin=1cm]{caption}

\numberwithin{equation}{section}	     

\title{Accelerated Computations of  Sensitivities for xVA}

\begin{document}

\author[1]{Griselda Deelstra}
\ead{Griselda.Deelstra@ulb.be}
\author[2,3]{Lech A.\ Grzelak}
\ead{L.A.Grzelak@uu.nl}
\author[1]{Felix L.\ Wolf\corref{cor1}}\
\ead{Felix.Wolf@ulb.be}
\cortext[cor1]{Corresponding author}
\address[1]{Department of Mathematics, Universit\'e libre de Bruxelles, Brussels, Belgium}
\address[2]{Mathematical Institute, Utrecht University, Utrecht, the Netherlands}
\address[3]{Rabobank, Utrecht, the Netherlands}

\footnotesize{
\begin{abstract}
\noindent
Exposure simulations are fundamental to many xVA calculations and are a nested expectation problem where repeated portfolio valuations create a significant computational expense. 
Sensitivity calculations which require shocked and unshocked valuations in bump-and-revalue schemes exacerbate the computational load. 
A known reduction of the portfolio valuation cost is understood to be found in polynomial approximations, which we apply in this article to interest rate sensitivities of expected exposures.
We consider a method based on the approximation of the shocked and unshocked valuation functions, as well as a novel approach in which the difference between these functions is approximated.
Convergence results are shown, and we study the choice of interpolation nodes. Numerical experiments with interest rate derivatives are conducted to demonstrate the high accuracy and remarkable computational cost reduction. We further illustrate how the method can be extended to more general xVA models using the example of CVA with wrong-way risk.

\end{abstract}
}
\begin{keyword}
\footnotesize{{xVA}, {Sensitivity},  {Exposure Simulation}, {Portfolio Approximation}, {Polynomial Interpolation}, {Quadrature Nodes}, {Monte Carlo}}
\end{keyword}
\maketitle


{\let\thefootnote\relax\footnotetext{The views expressed in this paper are the personal views of the authors and do not necessarily reflect the views or policies of their current or past employers.}}

%
\normalsize
\section{Introduction}
Value adjustments (xVA) are a core component in the risk management framework of modern asset pricing and are used to account for additional risks not covered under the standard theory of asset pricing. 
All xVA computations depend on the future evolution of the portfolios for which they are computed, so accurate exposure calculations are of high importance. 
{In the presence of} many stochastically modelled risk factors with potentially intricate dependencies, the exposure distribution quickly becomes intangible and warrants the use of Monte Carlo simulations.
With a large number of simulation paths and evaluation times of the exposures, an enormous amount of required portfolio valuations may aggregate, which acts as a significant driver behind the often large computational demand of the exposure simulation.
\par
One way to achieve faster simulations is through a reduction of the number of Monte Carlo paths or monitoring dates, but this is to the immediate detriment of accuracy. An alternative {approach is to improve} the speed of portfolio valuations.
\par
Approximation techniques provide such an opportunity to reduce the computational complexity along the number of Monte Carlo simulations. At every time step, instead of evaluating the exact, and expensive, portfolio valuation function directly in every realisation of its underlying risk factors, an accurate approximation is constructed from only a few exact evaluations. 
The resulting approximation can be evaluated very efficiently thanks to its polynomial structure.
{Similar approaches have been applied to improve efficiency in the simulation of differential equations \citep{XiuHighOrderCollocation} where they are known as stochastic collocation techniques. 
The approach considered in this article is closely connected to polynomial asset valuation approximation techniques, which have recently gained increased attention in the Mathematical Finance literature.}
{\cite{Ga2018} and \cite{Glau2019b} utilized polynomial interpolation over Chebyshev nodes for the pricing of options, respectively for the approximation of implied volatility functions. \cite{Glau2019a} introduced a novel approach to the pricing of American options, which uses Chebyshev interpolation to approximate the continuation values of the option.  This approach allows for the splitting of the computation into a construction phase, where the approximation functions are computed, and a numerically efficient evaluation phase.  In \cite{Glau2020}, this dynamic approach is extended to a broader class of options, and to the nested expectation problem which is posed by exposure simulations. Here, it is also remarked that approximation techniques could prove suitable for sensitivity computations.}
\par
{\cite{GrzelakColl} studied stochastic collocation techniques to improve the efficiency in sampling from random distributions, and showed optimality of quadrature interpolation nodes for (normally distributed) interest rates with respect to the $L^2$-norm. In \cite{grzelak2022sparse} the approximation of portfolio valuations with sparse grids in a multi-dimensional setting is considered.}
\par
{Here}, we extend the stochastic-collocation-based approach of exposure simulation to the calculation of {expected exposure sensitivities} at high numerical efficiency. 
We consider the calculation of interest-rate sensitivities, in particular the sensitivity with respect to market quotes used in the construction of the yield curve.  
These sensitivities are typically obtained from difference quotients, where the valuation under a particular {market shock} is compared to the valuation under normal, {unshocked} market conditions, so-called bump-and-revalue schemes {(see, for example,  \cite{green2015xva})}.%
\footnote{{In this article, ``sensitivity'' refers to the effect on the expected exposure induced by small changes to the term structure of interest rates. It can be thought of as a particular type of derivative, which is approximated by difference quotients. Then, ``shocked'' and ``unshocked'' valuations refer to the function valuations with and without shifted parameters, respectively. A comprehensive description of the procedure is provided in \Cref{sec:yieldcurvesensi}.}}
 For each market shock, such a scheme demands the full amount of exact portfolio valuations, which effectively multiplies the numerical effort by the number of market shocks considered.
\par
Many attempts to manage this complexity are based on novel techniques, such as Adjoint Algorithmic Differentiation \citep{Capriotti2017, Huge2017} or neural-network-based xVA engines \citep{Gnoatto2020}. 
However, such methods require substantial overhead in their implementation, {whereas} the approach introduced in this article can be seamlessly integrated into standard Monte Carlo exposure simulations and merely requires the implementation of a polynomial interpolation scheme. 
\par
{
In this article, we introduce a new method for constructing estimations of the interest rate sensitivity of a portfolio's expected exposure.
We demonstrate how a polynomial interpolation technique can be directly applied to the unshocked and shocked portfolio valuation functions, leading to a significant reduction in computational requirements for sensitivity calculations. Moreover, we introduce a more efficient approach in which the difference between shocked and unshocked valuation functions is approximated with a reduced-degree polynomial, leading to increased complexity reduction. We analyse convergence results and error bounds for the proposed methods and choices of interpolation nodes.
\par
Our proposed approximation technique is compatible with sophisticated xVA frameworks, and we outline an adaptation to models that include wrong-way risk.
\par
We conduct numerical experiments on portfolios of linear and non-linear assets, with a focus on large swap portfolios and Bermudan swaptions. In these experiments, we demonstrate how the $L^2$-optimal quadrature nodes automatically resolve the challenge of finding a suitable interpolation domain.
\par

}
%
%
%
%
\par
The present article is structured as follows. 
In \Cref{sec:exposuresimulation} we introduce exposure simulations in the classical sense and their approximations with polynomial interpolation. \Cref{sec:yieldcurvesensi} summarizes the concept of interest-rate sensitivities, i.e., sensitivities with respect to the market quotes used in the construction of the yield curve, and defines the sensitivity of expected exposures.
{
In \Cref{sec:ApproximatedEESensis}, two approximation methods for these sensitivities are introduced. A \emph{full-order approximation} applies the previously defined expected exposure approximations in a bump-and-revalue scheme. A \emph{low-order difference approximation} directly approximates the functional change under the market shock.%
}
\Cref{sec:errorana} {contains an analytical error analysis, where we show convergence of the sensitivity approximation and also discuss the selection of interpolation nodes.} 
In \Cref{sec:experiments} we {investigate the accuracy of the methods with numerical experiments, and in \Cref{sec:BermSwaptionExperiment} we detail how the approach can be applied to Bermudan Swaptions with an additional numerical experiment. Finally, we} summarize our findings in \Cref{sec:conclusion}.
%
%
%
\section{Exposure simulation and approximation}\label{sec:exposuresimulation}
{Before we introduce the sensitivity approximation methods studied in this article}, we provide {the context in which they arise, given by} exposure simulations for xVA.
{We consider portfolios with underlying risk factors modelled by a stochastic process $(X(t))_{t\geq0}$ on a filtered probability space $(\Omega, \F, (\F_t)_{t\geq0}, \Q)$, where $\Q$ is the risk-neutral measure. 
The stochastic process $X(t)$ takes values in $\R^m$, where $m$ corresponds to the number of modelled risk factors.
The portfolio value at time $t$ is then described by a stochastic process $(V(t, X(t)))_{t\geq 0}$. When all future cash flows can be expressed in terms of payoffs $H(T_k, X(T_k))$, $k\in\{1,\dots, L\}$, $L\in\mathbb{N}$, the portfolio valuation is given by the conditional expectation\footnote{We denote expectations conditional on the filtration $\F_t$ with a subscript $\E_t$.}}
\begin{equation}
V(t, X(t)) = \E_t^\Q\left[  \sum_{k=1}^L \frac{B(t)}{B(T_k)} H(T_k, X(T_k))\right], 
\end{equation}
where  $B(t) = \exp(\int_0^t r(s)\d s)$ is the num\'eraire of {the measure $\Q$, i.e.\ we denote the risk-neutral interest (short) rate by $r(t)$.}
\par
In many xVA models, the valuation adjustment depends on whether a portfolio is associated with a cost or a benefit, that is, the models depend on the sign of the (future) valuations. This is expressed by the notion of positive and negative exposures, which are given by the positive and negative part of the valuation functions,
$V^+(t, X(t))$ and $V^-(t, X(t))$, respectively.
Exemplarily for credit valuation adjustments, a default of the counterparty only affects those portfolios which have a positive present value at the time of default. Equivalently from the {point of} view of the counterparty, its own default event is associated with its debit valuation adjustment, which analogously is based on the negative exposure (as seen by the counterparty).
\par
{The deterministic} xVA quantities at {initial} time $t_0$ are based on \emph{expected (positive or negative) exposures}, which are defined as the expectations of the respective discounted exposures. The expected %
 positive%
\footnote{The expected negative exposure is identical to \eqref{eq:ee} with the positive exposure replaced by the negative exposure $V^-(t, X(t))$. Within this article, we consider only the positive exposure, by symmetry the results can be extended to negative exposure computations.} 
exposure {at monitoring date $t\geq t_0$ }
 is given by
\begin{equation}
\EE(t_0, t) = \E_{t_0}^\Q\left[ \frac{B(t_0)}{B(t)} V^+(t, X(t))\right]
\approx \frac1M \sum_{j=1}^M \frac{B(t_0)}{B(t;\omega_j)} V^+(t,X(t; \omega_j)), \label{eq:ee}
\end{equation}
{which} can be approximated by a Monte Carlo simulation approach with a large number {of simulations} $M\in\mathbb{N}$ of the underlying risk factors $X(t; \omega_j)$, $\omega_j \in \Omega$, $j\in\{1,\dots,M\}$.
\par
{
\begin{rem}
In this context, our focus is on the risk-neutral equivalent martingale measure $\Q$, which is suitable for simulating exposures for xVA pricing purposes. Nevertheless, the approach presented in this article is not limited to this measure and can also be used for computations under alternative measures such as the historical measure, or even technical choices such as the forward, swap, and other measures.
\end{rem}
}
\subsection{Construction of approximating portfolio valuation functions}\label{sec:portfolioappx}
Given $R\in\mathbb{N}$ monitoring dates and $M$ simulation paths, the Monte Carlo approximation in \eqref{eq:ee} requires $M\times  R$ valuations of the portfolio.
Each of these exact portfolio valuations may correspond to a contrived function evaluation, potentially due to the number of assets in the portfolio, or due to the presence of (non-linear) assets which require complex solution schemes. Therefore, the large number of exact portfolio valuations contributes an immense numerical expense to the exposure simulation.
However, at each monitoring date $t$, the shape of the portfolio valuation function $V(t, \cdot)$ often exhibits a high degree of regularity when seen as a surface $\R^m \to \R$. This suggests that alternative functions $g(t, \cdot)$ exist, which generate almost indistinguishable valuation surfaces but can be numerically evaluated at much higher efficiency. Replacing the exact portfolio valuations with such approximation functions drastically reduces the computational burden of exposure simulations.
\par
To obtain such approximating functions $g(t, \cdot)$ of the portfolio valuation functions $V(t, \cdot)$, we administer polynomial interpolation methods. 
The construction of a polynomial approximation based on $N$ interpolation points $\{(x_k(t), V(t, x_k(t))\colon k\in\{1,\dots,N\}\}$ requires only $N$ of the expensive, exact portfolio valuations, where $N$ is typically by orders of magnitude smaller than the number $M$ of exact portfolio valuations utilized in the classical Monte Carlo approach \eqref{eq:ee}. One of the challenges in the approach is to find appropriate interpolation nodes $x_k(t)\in\R^m$ in the domain of the underlying risk factor $X(t)$.
\par
For the remainder of the article, we will focus on the base case of $m=1$, i.e.\ {portfolios with a single, one-dimensional risk factor}. Particularly, in the numerical experiments of \Cref{sec:experiments} {and \Cref{sec:BermSwaptionExperiment}}, we will consider interest rate products whose price at time $t$ depends on the state of the underlying risk-free rate $r(t)$.
The approximating functions $g(t, \cdot)$ can then be explicitly given in the Lagrange form
\begin{equation}\label{eq:gt}
g(t, x) := \sum_{k=1}^N \left( V(t, x_k(t)) \prod_{\substack{\ell=1\\ \ell\neq k}}^N \frac{x-x_\ell(t)}{x_k(t)-x_\ell(t)} \right) \approx V(t, x),
\end{equation}
after a suitable choice of interpolation nodes {$(x_1(t), \dots, x_N(t))$ has been chosen, and the corresponding exact valuations $(V(t, x_1(t)), \dots, V(t, x_N(t)))$ have been computed. We define the set of $N$ interpolation nodes by $\bm x_N(t)$.}
The described polynomial of degree $N-1$ through a fixed choice of $N$ interpolation points is unique, {hence} a numerical implementation may equivalently utilize a different polynomial basis, or a numerically more stable form, like the barycentric interpolation formula. We refer to \cite{berrut2004barycentric} for a treatise on its derivation and advantages. 
{
\begin{rem}\label{rem:Vreg}
Equation \eqref{eq:gt} sheds light on the required level of regularity for the portfolio valuation V. Specifically, the Lagrange remainder $R(t, x) := V(t, x) - g(t, x)$ of the polynomial interpolation $g$ in $N$ points depends on the $N$th derivative of the target function. It holds, for some value $\xi \in [\min\limits_{1\leq k\leq N}(x_k), \max\limits_{1\leq k\leq N}(x_k)]$, that
\begin{equation}
R(t, x) = \frac{ \left. \frac{\d^N V(t, x)}{\d x^N}\right|_{x=\xi}}{N!} \prod\limits_{k=1}^N (x - x_k(t)).
\end{equation}
This suggests that the method may not be applicable if the $N$th derivative cannot be controlled. Conversely, a bounded $N$th derivative guarantee a certain quality of approximation over the domain spanned by the interpolation nodes.  In \Cref{sec:errorana}, we extend this notion for expected exposures to the entire domain $\R$ of the interest rate risk factor.
\end{rem}
}
\par
{Monte Carlo exposure simulations can be greatly simplified with a stochastic collocation method, where exact valuations $V(t, \cdot)$ in \eqref{eq:ee} are replaced by approximations $g(t, \cdot)$.}
This yields an approximation of the expected (positive) exposure.
\par
\begin{dfn}[Expected exposure approximation]\label{dfn:EEcoll}
At each monitoring date $t$, let $V(t, \cdot)\colon \R \to \R$ be a portfolio valuation function with stochastic risk factor $X(t)$. Let $g(t, \cdot)$ be an approximation of the portfolio valuation. Then, the expected (positive) exposure approximation is given by 
\begin{equation}
\widetilde \EE(t_0, t) = \E_{t_0}^\Q\left[ \frac{B(t_0)}{B(t)} g^+(t, X(t))\right] 
 \approx \frac1M \sum_{j=1}^M \frac{B(t_0)}{B(t;\omega_j)} g^+(t,X(t; \omega_j)), \label{eq:eecoll}
\end{equation}
with $M\in\mathbb{N}$ {samples of the risk factor.}
\end{dfn}
Notably, at every monitoring date $t$, each approximated expected exposure $\widetilde \EE(t_0, t)$ requires only $N$ (depending on the degree of the polynomials $g(t, \cdot)$) exact, expensive portfolio valuations and $M$ `cheap' valuations of the polynomial approximation, which can be undertaken at high numerical efficiency. This stands in contrast to the $M$ exact, expensive portfolio valuations in the classical approach \eqref{eq:ee}.
%
\begin{rem}
The approximation technique is not limited to basic xVA models that rely on isolated expected exposure terms. In general, the methods introduced here can be directly applied to many advanced xVA models, with the only requirement that explicit portfolio valuations take place within a Monte Carlo simulation. We outline the generalization on the example of unilateral CVA with wrong-way risk in \ref{sec:WWR}. 
\par
In the remainder of the article, we adopt the expected-exposure form {of \Cref{dfn:EEcoll}} {which suffices to demonstrate the proposed methodology and has a straightforward representation.}
\end{rem}
%
%
\section{Sensitivities with respect to the yield curve}\label{sec:yieldcurvesensi}
We {now} introduce the concept of sensitivity with respect to the yield curve and apply it to the expected exposures {previously defined}.
\par
In order to appropriately define the involved quantities and sensitivities, we briefly outline the yield curve construction. 
{The yield curve represents the term structure of interest rates, which makes it} a crucial tool in asset valuation. 
\par
\begin{dfn}[Yield curve and zero-coupon bond curve]
The yield curve $Y$ of an investment horizon $[t_0, T]$ 
{maps maturities to their associated total rate of return.}
That is, a payment of 1 (unit of the currency associated with the yield curve) at time $t_0$ returns an expected payment of $\exp(Y(S)(S-t_0))$ at time $S$. This is described by a real function
\begin{equation}
Y\colon [t_0, T] \to \R,\ S \mapsto Y(S). 
\end{equation}
Under no-arbitrage conditions, the yield curve is directly related to the zero-coupon bond curve $P(t_0, \cdot)$, which indicates the equivalent discounted value that a payment of 1 at time $S$ has at time $t_0$,
\begin{equation}\label{eq:discountcurve}
P(t_0, \cdot)\colon [t_0, T] \to \R^+, S \mapsto P(t_0, S) := \exp\left(-Y(S) (S-t_0)\right).
\end{equation}
\end{dfn}
A yield curve is calibrated to the market on the basis of a set of liquidly traded instruments with associated market quotes. We refer to this basis as the \emph{market instruments}. Yield curve calibration refers to the process of finding a curve that reproduces each of these market quotes when the model prices of the instruments are calculated with respect to the curve. 
Typically, each instrument will only contribute information towards {distinct parts} of the yield curve. To obtain the entire curve over a continuous spectrum, suitable curve-fitting schemes must be applied, {which affect} the resulting yield curve. 
{One such scheme is the multi-dimensional Newton--Raphson algorithm, which is applied in \cite{GrzelakOosterlee} to obtain a discrete set of points on the yield curve. The effect that various interpolation methods between such points have on the yield curve is studied in \cite{hagan2006interpolation}.}
\par
In this article, we use \emph{sensitivity to the yield curve} as a general moniker, also the term \emph{interest rate sensitivity} is known in the literature. To be precise, we refer to sensitivity with respect to the market quote of one of the market instruments {from which} the yield curve {is constructed}. 
{Sensitivity refers to the rate of change, in our considerations of the expected exposure, to a change in the aforementioned market quote. In other words, it is a difference quotient which approximates the derivative of the expected exposure with respect to the market quote considered. 
}
\par
{Each market instrument is associated with a particular maturity and particularly influences the yield curve in the surrounding region.}
Unlike approaches where only the finalized yield curve is shifted in a neighbourhood of the desired maturity, the sensitivity with respect to a market quote ensures that potential non-localities of the yield curve interpolation are correctly taken into account.
Since the effect of different market quotes are of interest, the sensitivity with respect to the yield curve is rather a family of sensitivities or a \emph{sensitivity profile}.
\par
 We formalise this in the following definitions.
%
	\begin{dfn}[Market instruments]\label{dfn:constructinginstruments}
	The \emph{market instruments} of a {yield curve $Y$} 
	are a set of market instruments 
	\begin{equation}
	\Phi = \{\varphi_1, \dots, \varphi_n\},
	\end{equation}
	where at time $t_0$, the market instrument $\varphi_i$, {$i \in \{1, \dots, n\}$}, corresponds to a maturity $T_i \in [t_0, T]$,
		 has an available market quote $K_i$
		and a valuation function in terms of the curve {$Y$}. 
	\end{dfn}
%
{In the following,} we define the sensitivity of the expected exposure of a portfolio in terms of difference quotients obtained from \emph{shocked} and \emph{unshocked} valuations. 
{Shocked valuations are the result of valuations using a shocked yield curve, which in turn is the yield curve obtained when a small shock, i.e.\ a linear shift, is applied to the quote of one of its market instruments.}
\begin{dfn}[Shocked yield curves]\label{def:shockedyieldcurve}
	Let $Y$ be a yield curve %
	obtained from the market instruments $\Phi$ with market quotes $\{K_1, \dots, K_n\}$. 
	For any $i\in\{1,\dots,n\}$, let $Y_i$ be the yield curve obtained from the same set of instruments $\Phi$, where the $i$-th instrument has its market quote shifted by a value $\Delta K_i \in \R$. That is, the market quotes of yield curve $Y_i$'s market instruments are $\{K_1, \dots, K_{i-1}, K_i + \Delta K_i, K_{i+1}, \dots, K_n\}$.
\end{dfn}	
We can now define the precise sensitivities considered in this article.
\begin{dfn}[Sensitivity of the expected exposure with respect to the yield curve]\label{def:yieldcurvesensi}	
	Let $V(t; Y)$ denote the portfolio valuation at time $t\in[t_0, T]$ under the {assumption of an initial} yield curve $Y$. For any $i\in\{1,\dots,n\}$, the sensitivity of $V$ with respect to $K_i$ is defined by
	\begin{equation}\label{eq:defdVdK}
	\frac{\partial V(t; Y)}{\partial K_i} := \lim\limits_{\Delta K_i \to 0} \frac{V(t; Y_i) - V(t; Y)}{\Delta K_i}.
	\end{equation}
	Let $\EE(t_0, t)$ be the expected (positive) exposure of $V$ as given in \eqref{eq:ee}. The sensitivity of the expected exposure at time $t$ with respect to the market quote $K_i$ is given by
	\begin{equation}\label{eq:defdEEdK}
	\frac{\partial \EE(t_0, t)}{\partial K_i} = \frac{\partial}{\partial K_i} \E_{t_0}^\Q\left[ \frac{B(t_0)}{B(t)} V^+(t, X(t)) \right].
	\end{equation}
\end{dfn}
The sensitivity of the expected exposure can be rewritten by exchanging the order of the expectation and differentiation operator, followed by the product rule of differentiation. This yields the formula
\begin{align}\label{eq:prodrule}
\frac{\partial \EE(t_0, t)}{\partial K_i}  &=  \E_{t_0}^\Q \left[ \frac{\partial}{\partial K_i} \left( \frac{B(t_0)}{B(t)} V^+(t, X(t)) \right)\right] \nonumber \\
&= \E_{t_0}^\Q\left[  \left( \frac{\partial}{\partial K_i} \frac{B(t_0)}{B(t)}\right) V^+(t, X(t)) + \frac{B(t_0)}{B(t)} \frac{\partial V^+(t, X(t))}{\partial K_i} \right]. 
\end{align}
This expanded form is suitable to approximation in the common framework of Monte Carlo exposure simulations and allows us to relate the sensitivity of the expected exposure to sensitivities of the num\'eraire and of the portfolio. 
\par
The derivatives %
in \eqref{eq:prodrule} {can not be assumed to be} analytically available, depending on the construction methodology of the yield curve and the type of assets in the portfolio. 
{Therefore,} we will consider forward differences for some fixed, small values of $\Delta K_i > 0$ in \eqref{eq:defdVdK}.
\par
We introduce some short-hand notations. 
In the unshocked market, we continue denoting the interest rate by $r(t)$, the money market account by $B(t)$, the risk factor by $X(t)$ and the value of the portfolio by $V(t, X(t))$. In the $i$-th shocked market, we denote the interest rate by $r_i(t)$, the money market account by $B_i(t) = \exp(\int_{0}^t r_i(s)\d s)$, the risk factor by $X_i(t)$ and the portfolio valuation by $V_i(t, X_i(t))$.
We emphasize that the market shock affects the entire discounting structure. 
\par
The resulting Monte Carlo simulation scheme for the sensitivity of an interest rate asset with $M$ paths $\omega_j$, $j=1,\dots,M$, is given by
\begin{multline}\label{eq:MCexactprod}
\frac{\partial}{\partial K_i} \EE(t_0, t) 
\approx 
\sum_{j=1}^M  \left( \frac{\partial}{\partial K_i} \frac{B(t_0; \omega_j)}{B(t; \omega_j)}\right) V^+(t, X(t;\omega_j)) + \frac{B(t_0; \omega_j)}{B(t; \omega_j)} \frac{\partial V^+(t, X(t;\omega_j))}{\partial K_i} \\
\approx 
\sum_{j=1}^M \frac{\frac{B_i(t_0; \omega_j)}{B_i(t; \omega_j)} - \frac{B(t_0; \omega_j)}{B(t; \omega_j)}}{\Delta K_i} V^+(t, X(t;\omega_j)) + \frac{B(t_0; \omega_j)}{B(t; \omega_j)} \frac{V_i^+(t, X_i(t;\omega_j)) - V^+(t, X(t;\omega_j))}{\Delta K_i}.
\end{multline}
This is equivalent to the forward sensitivity approach given in \cite{green2015xva}.
\section{Approximation of expected exposure sensitivities}\label{sec:ApproximatedEESensis}
Analogous to the expected exposure approximation in \Cref{dfn:EEcoll}, we define two approximations of the expected exposure sensitivity \eqref{eq:prodrule}. 
First, a \emph{full-order approximation}, in which both the shocked and unshocked portfolio valuation functions, $V(t, \cdot)$ and $V_i(t, \cdot)$ are approximated with polynomials of full degree $N-1$, i.e.\ constructed with $N$ exact portfolio valuations each.
Secondly, we define a \emph{low-order difference approximation}, where only $V(t, \cdot)$ is approximated at full degree, and the approximations of $V_i(t, \cdot)$ are constructed with a lower degree approximation of the difference between the shocked and unshocked portfolio valuation functions.
\begin{dfn}[The full-order approximation of the sensitivity]\label{dfn:fullorder}
For any monitoring date $t_0 \leq t \leq T$, let the exact portfolio valuation $V(t, \cdot)$ associated with the yield curve $Y$ be approximated by the polynomial $g(t, \cdot)$, based on $N$ interpolation nodes $\bm{x}_N(t)$, {given in \eqref{eq:gt}}.
\par
For a fixed $i\in\{1,\dots,n\}$, let $V_i(t, \cdot)$ be the exact portfolio valuation function associated with the yield curve $Y_i$ of the $i$-th market shock, given in \Cref{def:shockedyieldcurve}. Let $g_i(t, \cdot)$ be a polynomial approximation function of $V_i(t, \cdot)$ based on {$N$} interpolation nodes {$(x^i_1(t), \dots, x^i_N(t))$, the set of which we denote by $\bm x^i_N(t)$.}
\par
The {full-order approximation} of the sensitivity of the expected exposure with respect to market quote $K_i$, {$\Psi^i_N(t) \approx {\partial \EE(t_0, t)}/{\partial K_i}$}, is given by 
\begin{multline*}
\Psi^i_N(t) := \E_{t_0}^\Q\left[   \left(\frac{\partial}{\partial K_i}\frac{B(t_0)}{B(t)}\right)	 g^+(t, X(t)) + \frac{B(t_0)}{B(t)} \frac{g_i^+(t, X_i(t)) - g^+(t, X(t))}{\Delta K_i}    \right] \\
 \approx \sum_{j=1}^M \frac{\frac{B_i(t_0; \omega_j)}{B_i(t; \omega_j)} - \frac{B(t_0; \omega_j)}{B(t; \omega_j)}}{\Delta K_i} g^+(t, X(t;\omega_j)) + \frac{B(t_0; \omega_j)}{B(t; \omega_j)} \frac{g_i^+(t, X_i(t;\omega_j)) - g^+(t, X(t;\omega_j))}{\Delta K_i}.
\end{multline*}
\end{dfn}
{
\begin{rem}
There are two sources of approximation error in the above definition. First, the accuracy of the approximation $g$ hinges on the regularity of the portfolio valuation function $V$, as detailed in \Cref{rem:Vreg}. This approximation error may potentially be addressed by increasing the number of interpolation nodes $N$.
\par
Secondly, the construction of the yield curve itself may introduce inaccuracies. In \ref{appx:yieldcurve}, we show connections between the sensitivity of the yield curve and its modelling methodology, which in turn may affect the quality of the approximation. This source of inaccuracy requires refinement of the yield curve construction methodology and is not directly related to the interpolation of the portfolio valuation function.
\end{rem}}
\par
{The full-order approximation} greatly reduces the number of exact portfolio valuations. In the classical approach given in \eqref{eq:MCexactprod}, $2M$ exact portfolio valuations are required, where {we recall that} $M$ is the number of paths in the Monte Carlo simulation. In the full-order approach, this is reduced to $2N$ exact valuations to obtain the {approximation functions}, and $2M$ (cheap) valuations of the polynomial approximations.
%
The full-order approach to sensitivity in \Cref{dfn:fullorder} relies on polynomial approximations based on $N$ exact valuations at their interpolation nodes. In the computation of a sensitivity profile with respect to multiple market quotes $\{K_1, \dots, K_n\}$, each sensitivity approximation reuses the same approximation $g(t, \cdot)$ of $V(t, \cdot)$ but requires a newly constructed approximation $g_i(t, \cdot)$ of $V_i(t, \cdot)$. Thus, $n$ different market shocks require $(n+1)N$ exact portfolio valuations per monitoring date.
\par
We {now introduce a} low-order difference approximation which reduces the number of exact valuations needed to approximate the shocked portfolio valuation $V_i(t, \cdot)$. 
\par
In the low-order difference approach, for every monitoring date $t$, an approximation $\widetilde g_i(t, \cdot) \approx V_i(t, \cdot)$ is constructed based on the previously established estimator of the unshocked portfolio function $g(t, \cdot)$, to which another polynomial $h_i(t, \cdot)$ of lower degree is added.
That is, we set
\begin{equation}\label{eq:tildegi}
\widetilde g_i(t, \cdot) := g(t, \cdot) + h_i(t, \cdot).
\end{equation}
By constructing $h_i(t, \cdot)$ with $d<N$ interpolation nodes, only $d$ exact portfolio valuations $V^i(t, \cdot)$ are required, and consequently, the total number of exact portfolio valuations in the sensitivity profile is reduced to $nd+N$ at each monitoring date.
\par
We construct $h_i(t, \cdot)$ as a polynomial {of degree $d-1$} which approximates the difference between the polynomial approximation $g(t, \cdot)$ and the shocked portfolio, $V_i(t, \cdot)$. 
That is, in Lagrange form, the difference approximation $h_i(t, \cdot)$ is given by
\begin{align}\label{eq:htidef1}
h_i(t, x) &:= \sum\limits_{j=1}^d \left( V_i(t, x_{j,d}^i(t)) - g(x_{j,d}^i(t)) \right) L^i_{j,d,t}(x), 
\end{align}
{with Lagrange basis}
\begin{equation}
L^i_{j,d,t}(x) := \prod\limits_{\substack{k=1\\ k\neq j}}^d \frac{x - x_{k,d}^i(t)}{x_{j,d}^i(t) - x_{k,d}^i(t)},
\end{equation}
where {$(x^i_{1,d}(t), \dots, x^i_{d,d}(t))$ are the interpolation nodes used in the construction of $h_i(t, \cdot)$ which we collect in the set $\bm{x}^i_d(t)$.}
\par
We {now} define the low-order difference approximation of the sensitivity.
\begin{dfn}[The low-order difference approximation of the sensitivity]\label{dfn:lowdegreedifference}
For any monitoring date $t_0 \leq t \leq T$ and shock to market quote $K_i$, $i\in\{1,\dots,n\}$,
let the unshocked portfolio valuation function $V(t, \cdot)$, its approximation $g(t, \cdot)$, and the shocked portfolio valuation function $V_i(t, \cdot)$ be defined as in \Cref{dfn:fullorder}. Let $\widetilde g_i(t, \cdot)$ be the approximation of $V_i(t, \cdot)$ given in \eqref{eq:tildegi} and \eqref{eq:htidef1}.
The low-order difference approximation of the sensitivity of the expected exposure with respect to market quote $K_i$, $\Psi^i_{d,N}(t) \approx {\partial \EE(t_0, t)}/{\partial K_i}$, is given by 
\begin{equation}
\Psi^i_{d,N}(t) := \E_{t_0}^\Q\left[   \left(\frac{\partial}{\partial K_i}\frac{B(t_0)}{B(t)}\right) g^+(t, X(t)) + \frac{B(t_0)}{B(t)} \frac{\widetilde g_i^+(t, X_i(t)) - g^+(t, X(t))}{\Delta K_i}    \right].
\end{equation}
\end{dfn}
\par
When the {set of } interpolation nodes $\bm{x}^i_d(t)$ %
of $h_i(t, \cdot)$ are chosen as a subset of the $N$ interpolation nodes $\bm{x}^i_N(t)$ %
on which the approximation $g_i(t, \cdot)$ in the full-order approach is based, another interpretation of the low-order difference method arises.
Under this particular choice of interpolation nodes, it {becomes} possible to rewrite \eqref{eq:htidef1} as
\begin{align}\label{eq:htidef2}
h_i(t, x) &=  \sum\limits_{j=1}^d V_i(t, x_{j,d}^i(t))  L^i_{j,d,t}(x) -  \sum\limits_{j=1}^d g(t, x_{j,d}^i(t)) L^i_{j,d,t}(x) \nonumber \\
&= \sum\limits_{j=1}^d g_i(t, x_{j,d}^i(t))  L^i_{j,d,t}(x) -  \sum\limits_{j=1}^d g(t, x_{j,d}^i(t)) L^i_{j,d,t}(x),
\end{align}
since $g_i$ and $V_i$ coincide in the interpolation nodes $x_{j,d}^i(t)$.
{Thus, we may} relate $h_i(t)$ solely to the approximations $g_i$ and $g$. In particular, we may interpret $h_t^i$ as the difference between a polynomial approximation $p_i(t, \cdot)$ of $g_i(t, \cdot)$, and a polynomial approximation $p(t, \cdot)$ of $g(t, \cdot)$, both of reduced degree $d-1$. 
Summarizing, 
\begin{equation}\label{eq:hti-pip}
h_i(t, x)  =: p_i(t, x) - p(t, x).
\end{equation}
\par
With this representation at hand, it becomes easy to see that if the difference approximation is constructed with the full number of $d=N$ points, then, by the uniqueness of the polynomial approximation, it holds that $p_i(t, \cdot) = g_i(t, \cdot)$ and $p(t, \cdot) = g(t, \cdot)$. Consequently, the approximations $\widetilde g_i(t, \cdot)$ and $g_i(t, \cdot)$ must coincide and the same holds for the expected exposure sensitivity estimators $\Psi^i_{d,N}(t)$ and $\Psi^i_N(t)$, {when $d=N$}.
\section{Error Analysis}\label{sec:errorana}
We have thus introduced two approximators of the expected exposure's yield curve sensitivity, the full-order estimator $\Psi_N^i(t)$ given in \Cref{dfn:fullorder} and the low-order difference estimator $\Psi^i_{d,N}(t)$ given in \Cref{dfn:lowdegreedifference}.
The benchmark of the error analysis is provided by the conventional estimator, the finite difference approach with exact portfolio valuations given by
\begin{equation}\label{eq:Psi}
\Psi^i(t) := \E_{t_0}\left[ \left(\frac{\partial }{\partial K_i}\frac{B(t_0)}{B(t)}\right) V^+(t, X(t)) + \frac{B(t_0)}{B(t)} \frac{V_i^+(t, X(t)) - V^+(t, X(t))}{\Delta K_i} \right].
\end{equation}
\par
{For every time $t$, we analyse the absolute approximation error between the conventional estimator and the low-order difference approximation. Since it holds that $\Psi^i_{d,N}(t) = \Psi_N^i(t)$ for the special case of $d=N$, the resulting error bound can be quickly extended to the full-order approach.}
The absolute error is given by
\begin{multline}\label{eq:Psierror1}
|\Psi^i(t) - \Psi^i_{d,N}(t)|
 = \Biggl| \E_{t_0} \Biggl[ \left(\frac{\partial }{\partial K_i}\frac{B(t_0)}{B(t)}\right) \left(V^+(t, X(t)) - g^+(t, X(t))\right) \nonumber \\ 
 \qquad + \frac{B(t_0)}{B(t)} \left( \frac{V_i^+(t, X_i(t)) - V^+(t, X(t))}{\Delta K_i} - \frac{\widetilde g_i^+(t, X_i(t)) - g^+(t, X(t))}{\Delta K_i}\right) \Biggr] \Biggr|.
\end{multline}
In the following, {we will not write the second argument of the functions $V, V_i, g, g_i$ when using} $L^q(\Omega)$ norms for $q\in\{1, 2\}$. 
Observe that for any two arbitrary functions $f_1, f_2$, it holds that
\begin{equation}
| f_1^+(x) - f_2^+(x)| \leq |f_1(x) - f_2(x)|.
\end{equation}
Together with an application of Jensen's inequality and the triangle inequality, we obtain a first upper bound
\begin{align}
&|\Psi^i(t) - \Psi^i_{d,N}(t)| 
 \leq \E_{t_0} \left[ \left| \left( \frac{\partial }{\partial K_i}\frac{B(t_0)}{B(t)}\right) (V(t, X(t)) - g(t, X(t)))\right| \right] \nonumber \\
&\hspace{24pt}
+ \E_{t_0} \left[ \left|\frac{B(t_0)}{B(t)} \frac{ V_i(t, X_i(t)) - \widetilde g_i(t, X_i(t))}{\Delta K_i}\right|\right] + \E_{t_0} \left[\left| \frac{B(t_0)}{B(t)}\frac{V(t, X(t)) - g(t, X(t))}{\Delta K_i}\right| \right] \nonumber \\
& = \left\| \left( \frac{\partial }{\partial K_i}\frac{B(t_0)}{B(t)}\right) (V(t) - g(t)) \right\|_1 + \left\| \frac{B(t_0)}{B(t)} \frac{ V_i(t) - \widetilde g_i(t)}{\Delta K_i}\right\|_1 + \left\| \frac{B(t_0)}{B(t)}\frac{ V(t) - g(t) }{\Delta K_i} \right\|_1. \nonumber 
\end{align}
An application of  H{\"o}lder's inequality allows us to relate the absolute error directly to distances between target functions and their approximations,
\begin{align}
|\Psi^i(t) - \Psi^i_{d,N}(t)| 
&\leq \left\| \left( \frac{\partial }{\partial K_i}\frac{B(t_0)}{B(t)}\right)\right\|_2 \left\|V(t) - g(t) \right\|_2 \nonumber  \\
&\quad + \left| \frac{1}{\Delta K_i}\right| \left\| \frac{B(t_0)}{B(t)}\right\|_2 \left(\left\|V_i(t) - \widetilde g_i(t) \right\|_2 + \left\|V(t) - g(t)\right\|_2\right). \nonumber 
\end{align}
\par
The terms related to the discount factor in this inequality are quickly bounded. 
Because of its connection to the zero-coupon bond price $P(t_0, t)$, we immediately obtain that the $L^2$-norm of the discount factor is finite, and denote it by
\begin{equation}\label{eq:C1}
C_1(t) := \left\| \frac{B(t_0)}{B(t)}\right\|_2 = \E_{t_0}\left[  \exp\left(-2\int_{t_0}^t r(s)\d s\right) \right]^{\frac12} <\infty.
\end{equation}
A bound on the norm of the derivative of the discount factor is encoded in the assumptions of the yield curve construction. Generally, it holds that
\begin{align}\label{eq:C2}
C_2(t) := \left\| \frac{\partial }{\partial K_i}\frac{B(t_0)}{B(t)} \right\|_2 
&= \E_{t_0}\left[ \left( \frac{\partial }{\partial  K_i} \exp\left(-\int_{t_0}^t r(s)\d s\right) \right)^2 \right]^{\frac12} \nonumber \\
&= \E_{t_0}\left[ \left(\int_{t_0}^t \frac{\partial }{\partial K_i} r(s)\d s\right)^2  \exp\left(-2\int_{t_0}^t r(s)\d s\right)   \right]^{\frac12}.
\end{align}
One of the desirable features of yield curve interpolation described by \cite{hagan2006interpolation} is \emph{(forward) stability}, which describes how much the interpolated yield changes with respect to the inputs. The authors empirically confirm boundedness for a broad range of interpolation methods. Connecting forward stability to the interest rate becomes possible under concrete assumptions about the interest rate model used. 
{In \ref{sec:appx1}, this is shown explicitly for the 1-factor Hull--White interest rate model.}
{Thus, for} suitable interest rate and yield curve models, we may assume a bound 
\begin{equation}
|\Psi^i(t) - \Psi^i_{d,N}(t)| 
\leq C_2(t) \left\|V(t) - g(t) \right\|_2 
 + \frac{C_1(t)}{|\Delta K_i|}\left(\left\|V_i(t) - \widetilde g_i(t) \right\|_2 + \left\|V(t) - g(t)\right\|_2\right), \nonumber
\end{equation}
where $C_1(t), C_2(t) \in \R$ are the constants {in \eqref{eq:C1} and \eqref{eq:C2}}, independent of the number of nodes $N$ and $d$.
\par
In the following, we identify the case $i=0$ with the market without parameter shocks, so that $V_0(t, x) = V(t, x)$ and $g_0(t, x) = g(t, x)$. Then, the approximation errors of the portfolio valuation function, respectively its shocked market versions, are given by 
\begin{equation}\label{eq:epsilonVg}
\varepsilon_i(t) :=  \|V_i(t) - g_i(t)\|_2, \quad i \in \{0, \dots, N\}.
\end{equation}

Recall that $\bm{x}^i_N(t)$ are the $N$ interpolation nodes used in the construction of $g_i(t)$ in the full-order approach and $\bm{x}^i_d(t)$ are the $d$ interpolation nodes used in the construction of $\widetilde g_i(t)$ in the low-order difference approach. In the previous section, it was shown that choosing the interpolation nodes of the low-order difference approach as a subset of the nodes in the full-order approach, 
\begin{equation}\label{eq:nestedpts}
\bm{x}^i_d(t) \subseteq \bm{x}^i_N(t)\quad \text{ for all } d \leq N,
\end{equation}
allows for a convenient interpretation of the low-order difference polynomial $h_i(x)$. 
It was shown in \eqref{eq:hti-pip}, that the low-order difference polynomial can be expressed by $h_i(t, \cdot) = p_i(t, \cdot) - p(t, \cdot)$, a difference between the degree $d-1$ approximations of $g_i(t, \cdot)$ and $g(t, \cdot)$.
This allows for further bounds
\begin{align}\label{eq:Vi-tildegi}
\| V_i(t) - \widetilde g_i(t) \|_2 
&\leq \| V_i(t) - g_i(t)\|_2 + \|g_i(t) - \widetilde g_i(t)\|_2 \nonumber \\
&= \varepsilon_i(t) + \|g_i(t) -  g(t) - (p_i(t) - p(t))  \|_2 \nonumber \\
&\leq \varepsilon_i(t) + \| g_i(t) - p_i(t) \|_2  +  \| g(t) - p(t) \|_2.
\end{align}
Analogously to \eqref{eq:epsilonVg}, we set $p_0(t) := p(t)$ and denote the errors of approximating the degree $N-1$ polynomials with their degree $d-1$ counterparts by
\begin{equation}\label{eq:deltapg}
\delta_i(t) := \| g_i(t) - p_i(t) \|_2, \qquad i \in \{0, \dots, N\},
\end{equation}
so that, combining these results, we obtain {the error bound of the low-degree approximation,}
\begin{equation}\label{eq:Psierrorfull}
|\Psi^i(t) - \Psi^i_{d,N}(t)| \leq C_2(t) \varepsilon_0(t) 
+  \frac{C_1(t)}{|\Delta K_i|} ( \varepsilon_0(t) + \varepsilon_i(t) + \delta_0(t) + \delta_i(t) ).
\end{equation}
{For $d = N$, it holds that $\delta_i(t) = 0$ and \eqref{eq:Psierrorfull} provides the error bound for the full-order approximation. Its convergence follows in the next section by showing that $\varepsilon_i \to 0$ as $N \to\infty$.}
%
%
\subsection{Quadrature nodes}\label{sec:quadraturenodes}
For $i\in\{0, \dots, n\}$, let $I^i_t:=[\min(\bm x^i_N(t)), \max(\bm x^i_N(t))]$ denote the interval spanned by the interpolation nodes $\bm x^i_N(t) = \{x^i_1(t), \dots, x^i_N(t)\}$, where we set $\bm x^0_N(t) := \bm x_N(t)$. {Furthermore, let $|I^i_t|$ denote its length.}
When the target function $V_i(t, x)$ is sufficiently regular, the accuracy of its polynomial interpolation $g_i(t, x)$ can be explicitly given, for every $x \in I^i_t$, in terms of the Lagrange bound,
\begin{equation}\label{eq:lagrange}
V_i(t, x) - g_i(t, x) =  \frac{ \left. \frac{\d^N V_i(t, x)}{\d x^N}\right|_{x=\xi_i}}{N!} \prod\limits_{k=1}^N (x - x^i_k(t)), \quad i \in \{0, \dots, n\},
\end{equation}
where the derivative of $V_i(t, x)$ is evaluated in some value $\xi_i \in I^i_t$. 
Since the approximation setting considered in this article does not permit control over the function $V_i(t, x)$ or its derivatives, we may only exert influence over the error of the interpolation through the choice of interpolation nodes $\bm x^i_N(t)$. 
The optimal choice of interpolation nodes depends on the norm under which $V_i(t, x) - g_i(t, x)$ is considered.
\par
The $L^2(\Omega)$ norm in \eqref{eq:epsilonVg} implicitly depends on the random variable $X_i(t)$, where we again set $X_0(t) := X(t)$. 
When the risk-factor $X_i(t)$ admits a density $f_{X_i(t)}$, the error with respect to the $L^2(\Omega)$ norm can be expressed as an integral with weight function $f_{X_i(t)}(x)$, and  \cite{GrzelakColl} showed that with respect to the $L^2(\Omega)$ norm required in \eqref{eq:epsilonVg}, the optimal choice of interpolation nodes are the quadrature points $y^i_k(t)$ of the associated quadrature rule,
\begin{align}%
\varepsilon_i(t)^2 &= \int_\R \left(V_i(t, x) - g_i(t, x)\right)^2 f_{X_i(t)}(x) \d x \nonumber \\
&= \sum_{k=1}^N \left(V_i(t, y^i_k(t)) - g_i(t, y^i_k(t))\right)^2 w^i_k(t) +\widetilde \varepsilon_N^i(t).
\end{align}
Here, $w^i_k(t)$ are the corresponding quadrature weights and $\widetilde \varepsilon^i_N(t)$ is the quadrature error. The quadrature points $y^i_k(t)$ can be constructed with the recurrence relation of \cite{golub1969calculation}, whenever the corresponding moments of $X_i(t)$ are available. Alternatively, when the orthogonal polynomial family of $L^2(\Omega)$ is known, they can be found as the zeros of these polynomials.
Choosing the interpolation points in this manner as $x_k^i(t) = y^i_k(t)$ yields
\begin{equation}
\varepsilon_i(t)^2 = \widetilde \varepsilon_N^i(t),
\end{equation}
and convergence $\varepsilon_i(t) \to 0 $, as $N \to \infty$, can be inferred from convergence of the respective quadrature rule, which depends both on the risk factor $X_i(t)$ and the valuation function $V_i(t)$. 
\par
In \Cref{sec:experiments} and \Cref{sec:BermSwaptionExperiment}, we will consider interest rate products with normally distributed risk factors $X(t) = r(t)$. Then, the optimal interpolation points are given by the zeros of the Hermite polynomials, and the convergence of the quadrature error, as $N \to \infty$, is obtained for sufficiently differentiable portfolio valuation functions $V_i(t)$ from the convergence of the Gauss-Hermite quadrature error \citep{abramowitz1964handbook}.
\par
Of the products considered, interest rate swap yield a valuation function which is a linear combination of analytic functions and is thus easily seen as sufficiently smooth. 
\cite{Ga2018} prove that the valuation function of European options is analytic, which can be applied to (European) swaptions. 
\cite{Glau2020} note that the valuation function of a Bermudan swaption is merely continuously differentiable, but also offer a practical method to determine convergence based on the decay of polynomial coefficients as the degree of the approximation function increases, a comparable approach is discussed in \cite{trefethen2019approximation}.
\par
The approximation errors of the low-order contribution, $\delta_i(t)$, $i\in\{0, \dots, N\}$, given in \eqref{eq:deltapg}, 
{can be bounded over their domain of interpolation with the Lagrange error \eqref{eq:lagrange},} where now the target functions $g_i(t)$ are polynomials and therefore analytic functions by construction. 
To ensure the nestedness of interpolation nodes $\bm x^i_d(t)$ for $d < N$, we choose them as the inner sets of $\bm x^i_N(t)$, where nodes are alternatingly removed from either end of the interval until the desired number of nodes remains,
\begin{equation}\label{eq:quadraturesubnodes}
\bm x^i_d(t) := \{x^i_{1 + \lfloor \frac{N-d}{2} \rfloor}(t), \dots, x^i_{N - \lceil \frac{N-d}2 \rceil} (t)\}.
\end{equation}
%
\subsection{Chebyshev nodes}
An alternative selection of interpolation nodes is given by the Chebyshev nodes. {These are given by $c_k := \cos(\frac{kx}{N})$ for the interval $[-1, 1]$, and can be scaled to the domain $I^i_t$ with a linear transformation $T^i_t\colon [-1, 1] \to I^i_t$.
It is well-known, see for example \cite{burden2015numerical}, that Chebyshev nodes allow for a stricter bound of the Lagrange error over their interpolation domain. From this, it can be deduced that with respect to the supremum norm it holds that
\begin{equation}
\|V_i(t, x) - g_i(t, x)\|_\infty \leq \frac{|I^i_t|^N}{2^{N} N!} \left\| \left. \frac{\d^N V_i(t, x)}{\d x^N}\right|_{x=\xi} \right\|_\infty.
\end{equation}
In this norm, the choice of Chebyshev interpolation nodes $x^i_k(t) = T^i_t(c_k)$ is considered optimal and convergence $\|V_i(t, x) - g_i(t, x)\|_\infty \to 0$ for $N \to \infty$ is easily shown whenever $V_i(t,x)$ is sufficiently regular.}
\par
The supremum norm is easily interpreted as the maximal deviation of the approximation function $g_i(t, x)$ from its target $V_i(t, x)$, and it implies convergence with respect to many other norms. %
However, these results are limited to the necessarily finite domain of interpolation $I^i_t$. 
%
%
\section{Numerical experiments: interest-rate swaps}\label{sec:experiments}
{In this section, we conduct numerical experiments on portfolios of interest rate swaps, i.e., the underlying risk factor of the portfolio is $X(t)=r(t)$. }
In the first experiment, we explore the behaviour of the proposed methods with a small portfolio comprised of a single swap.
Then, we move on to a larger portfolio, which consists of many different payer and receiver swaps. %
\par
In all experiments, we construct a yield curve based on the market instruments given in {\ref{sec:numerical-params}}, \Cref{tab:constructing-instruments}. Each of these instruments is associated with a market quote $K_i$ and a maturity $T_i$, %
$i\in\{1,\dots, 8\}$. 
Sensitivities with respect to the yield curve are based on individual shocks to each of these rates $K_i$, so that in total, $n=8$ additional, shocked yield curves are constructed as delineated in \Cref{def:shockedyieldcurve}. As we will see shortly, there exists a correspondence between the maturity $T_i$ of the shocked instrument, and the region of effect on the yield curve. 
This motivates a practise of referring to the shock of instrument $K_i$ as the shock at maturity $T_i$. 
\par
To {obtain a good approximation of the derivative \eqref{eq:prodrule},} we fix the shock size to $\Delta K_i = 0.0001 = 1$\,bp.
\par
The interest rates $r(t)$, and its shocked-market versions $r_i(t)$, are modelled with 1-factor Hull--White models \citep{hull1990pricing}. In the following, we continue to identify the case $i=0$ with the market without parameter shocks, so that $r_0(t) = r(t)$. The prescribed dynamics are 
\begin{equation}\label{eq:rt}
\d r_i(t) = \lambda (\theta_i(t) - r(t))\d t + \eta W(t), \quad 0\leq i\leq n,
\end{equation}
with a Brownian motion $W(t)$, speed of mean reversion $\lambda = 0.01$, volatility coefficient $\eta = 0.02$, and time-dependent drift term $\theta_i(t)$ chosen such that the bond prices obtained from the model fit the respective yield curves. Later on, in \Cref{sec:singleswap}, we also consider a stressed scenario with an increased volatility coefficient.
\par
At each time $t$, the interest rates follow normal distributions $r_i(t) \sim \mathcal{N}(\mu_i(t), \sigma_i^2(t))$ with explicitly known distribution parameters. As discussed in \Cref{sec:quadraturenodes}, we choose the interpolation nodes $\bm x^i_N(t)$ based on these distributions as the zeros of the (probabilist's) Hermite polynomials orthogonal under $L^2(\Omega)$, and the reduced interpolation nodes $\bm x^i_d(t)$ as their inner subsets given in \eqref{eq:quadraturesubnodes}.
%
%
\subsection{Single swap portfolio}\label{sec:singleswap}
We consider a portfolio consisting of a single interest-rate payer swap with twice-yearly payments at times $\bar T_1, \dots, \bar T_m$, final maturity $\bar T_m = 20$ years,  and a notional of $\bar N = 10000$ units of the currency. 
The fixed swap rate $\bar K$ is chosen as the par rate $\bar K = 0.02226$, i.e., it holds $V(t_0) = 0$ at initial time $t_0 = 0$.
\par
Initially, the number $N$ of interpolation nodes needs to be determined. 
{In practice, an approximation $g(t, \cdot)$ of the valuation function $V(t, \cdot)$ is already obtained during calculations of the expected exposure. This can be reused in the sensitivity approximations to the effect of improved numerical efficiency.}
{For this experiment}, we find that $N=7$ interpolation nodes are sufficient to achieve a relative approximation error of the expected exposure below 1\,bp,
\begin{equation}\label{eq:epsilonee}
\varepsilon_\EE := \left|\max\limits_{t\in[t_0, T]} \left(\frac{\widetilde \EE(t_0, t) - \EE(t_0, t)}{\EE(t_0, t)}  \right)\right| \approx 2.7 \times 10^{-5} < 1\, \text{bp}.
\end{equation}
\par
The output of the sensitivity calculations is a sensitivity profile of the expected exposure of the portfolio, which is a collection of all the sensitivities $\partial \EE(t_0, t)/\partial K_i$, for certain $K_i$ within the market instruments. In these numerical experiments, we consider sensitivity with respect to all market quotes used, $1 \leq i \leq 8$. 
\par
\begin{figure}[]
\centering
\includegraphics[width=.49\textwidth]{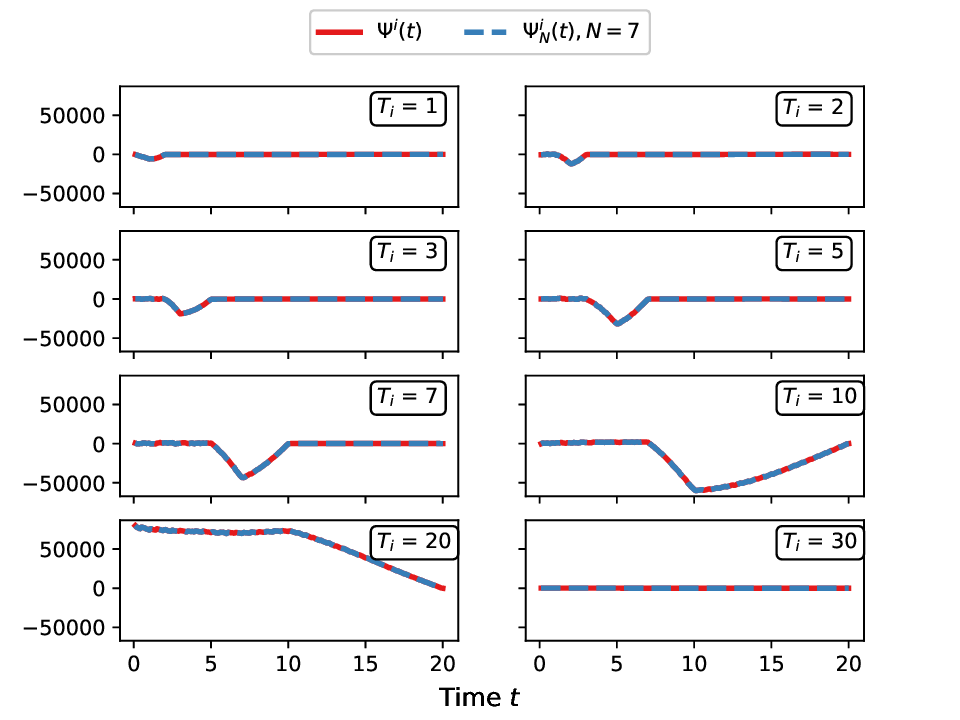}
\includegraphics[width=.49\textwidth]{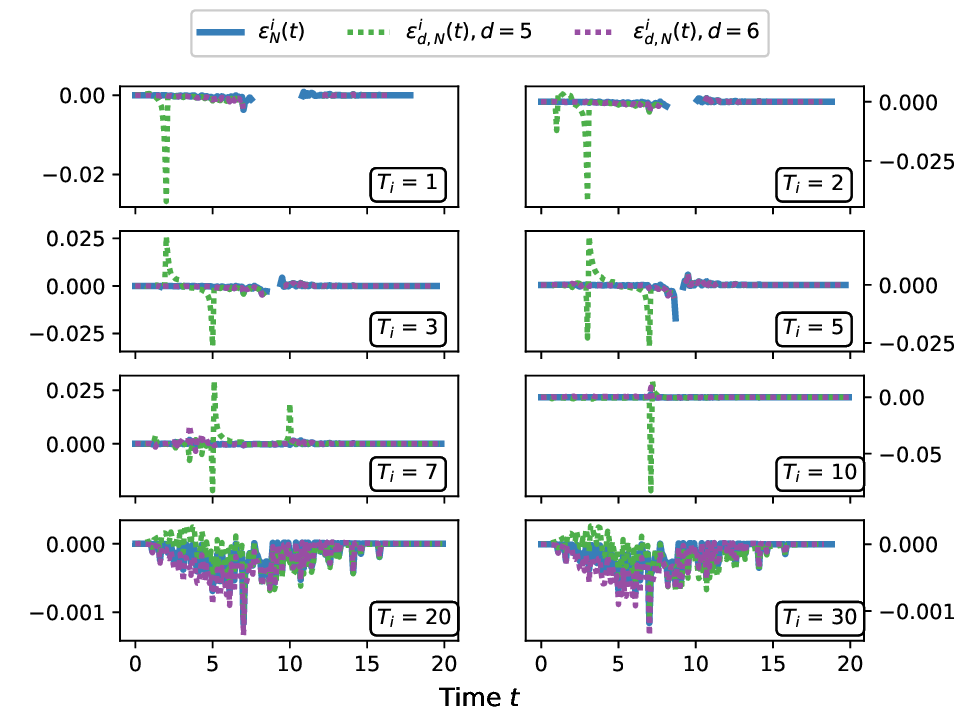}
\caption{
Left: Sensitivity profile obtained from exact approach $\Psi^i(t)$ and the full-order approximations $\Psi_N^i(t)$ for $1\leq i \leq 8$. 
Right: Relative approximation errors of the full-order approach $\varepsilon^i_N(t)$, %
and of the low-order difference approach, $\varepsilon^i_{d,N}(t)$.  $N=7$, $d\in \{5, 6\}$. %
{It should be noted that the relative approximation errors are negligible or confined to small peaks that have a minimal effect on the time integral of the sensitivity curve, which is ultimately utilized in xVA metrics.}
}
\label{fig:SwapSensi}
\end{figure}

In the collection of graphs on the left of \Cref{fig:SwapSensi}, the sensitivity profile of the swap portfolio is shown. 
It is computed both with the full-order estimator of the sensitivities,
$\Psi^i_N(t) \approx \frac{\partial \widetilde \EE(t_0, t)}{\partial K_i}$, defined in \Cref{dfn:fullorder}, and with the exact valuation approach 
$\Psi^i(t) \approx \frac{\partial \EE(t_0, t)}{\partial K_i}$, given in \eqref{eq:Psi}.
For each $i \in \{1, \dots, 8\}$, the market instruments $K_i$ correspond with maturities $T_i \in \{1, 2, 5, 7, 10, 20, 30\}$. We observe that the magnitude of the sensitivities increases for $i\in\{1,\dots,7\}$, up to the special quote $K_7$, whose maturity is equal to the maturity of the swap in the portfolio itself,  $T_7 = 20 = \bar T_m$. 
With respect to the market quote $K_8$ which is associated with the maturity $T_8 = 30$, there is nearly no sensitivity as the portfolio expires beforehand.
\par
%
%
%
%
%
In the graphical sensitivity profiles, no difference between the full-order approach and the exact valuations is discernible. To demonstrate the differences, we continue with representations of the approximation errors, and additionally include the low-order difference approximation $\Psi^i_{d,N}(t)$, defined in \Cref{dfn:lowdegreedifference}.
\par
For each of the sensitivities with respect to $K_i$, $1\leq i\leq 8$, we compute a relative approximation error of the full-order approach, and of the low-order difference approximations.  These relative errors are defined as the total approximation errors normalized with the exact result $\Psi_i(t)$, that is,
\begin{equation}\label{eq:relerror}
\varepsilon^i_N(t) := \frac{\Psi_N^i(t) - \Psi^i(t)}{\Psi^i(t)}, \quad
\varepsilon^i_{d, N}(t) := \frac{\Psi^i_{d,N}(t) - \Psi^i(t)}{\Psi^i(t)},
\end{equation}
respectively. The relative errors are not well-defined for $\Psi^i(t) = 0$ and only numerically stable, when $\Psi^i(t)$ is not too close to zero. In these cases, we refrain from computing the relative error.
\par
The approximation errors of the single-swap portfolio are given in the graphs to the right of \Cref{fig:SwapSensi}, computed for $N=7$ and for $d=5$, $d=6$.
At most times, all three considered methods display an extremely small relative approximation error below $0.1\%$. There are a few exceptions at individual time points, {however}, the {error} remains very small. For the low-order difference approaches, it can be found to be at most $7\%$ for $d=5$, $0.6\%$ for $d=6$, and $0.2\%$ for the full-order approach ($d=N$). 
In the context of computing xVA functions, the expected exposures are often integrated over time, which further minimizes the impact of these spike-type errors. This is further investigated in \Cref{sec:largeportfolio}.
\par
\par
The efficiency improvement provided by the method is highly promising. Recall that $M$ is the number of Monte Carlo paths which can easily range in the tens of thousands for exposure simulations.  Under the classical, exact valuation approach, it is necessary to compute $M$ exact valuations at each time point for the unshocked market, and another $M \times n$ exact valuations for the $n$ shocked markets considered. In this example, this attributes to $180000$ exact valuations at each monitoring time $t$.
\par
In the full-order approach with $N$ interpolation nodes, at each time $t$ only $(n+1) \times N$ exact valuations ($63$ in this example) are required. Additionally, the approximation function must be constructed, which attributes a numerical complexity of at worst $\mathcal{O}(N^2)$ per approximation function. Finally, the approximated portfolio valuations require $M$ {evaluations} of the approximation function, which are always cheap to evaluate, {regardless of} the choice of {the} exact valuation function. 
For the low-order difference method, the number of exact valuations is further reduced to $N + n \times d$ ($47$ and $55$, respectively, for $d=5$, $d=6$). This efficiency advantage increases further, as greater sensitivity profiles with more market shocks $n$ are considered. 
\par
We demonstrate the advantage of using the low-order difference approach over the more straightforward full-order approach with a reduced degree $N$.
To this end, we consider the low-order difference approach $\Psi^i_{d,N}(t)$ with $N=7$ and $d=5$, for which $47$ exact valuations are required. It is compared to the full-order approaches $\Psi^i_N(t)$ for $N=5$ and $N=6$, with $45$ and $54$ exact valuations, respectively. 
This choice provides a similar number of exact valuations and we will demonstrate that the baseline parameter $N=7$ in the low-order approach is not simply overspecified.
\par
\begin{figure}[]
\centering
\includegraphics[width=.49\textwidth]{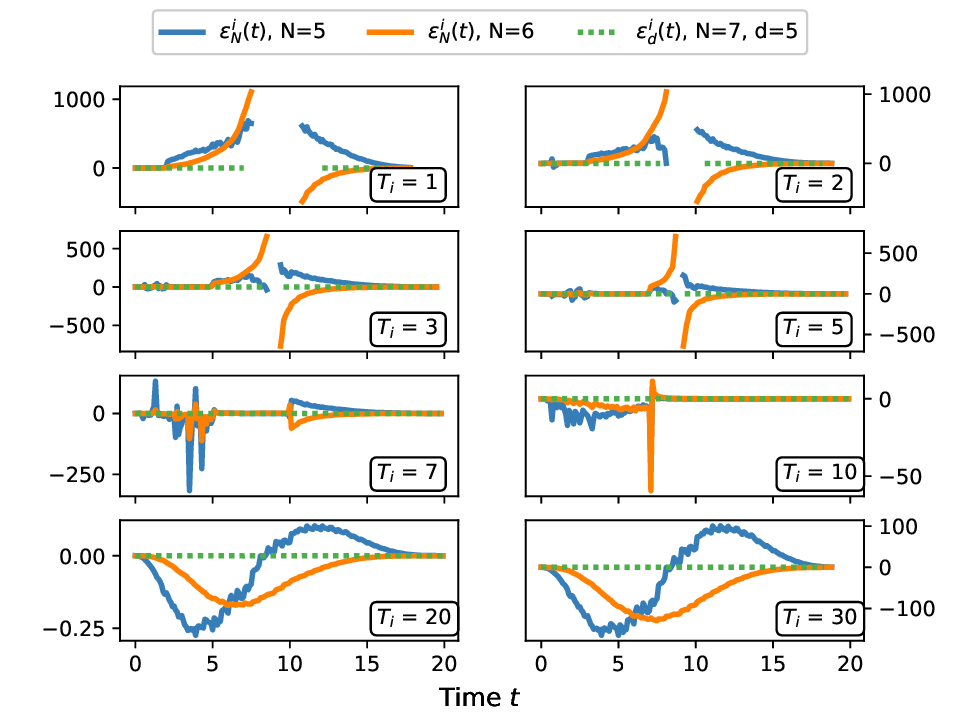}
\includegraphics[width=.49\textwidth]{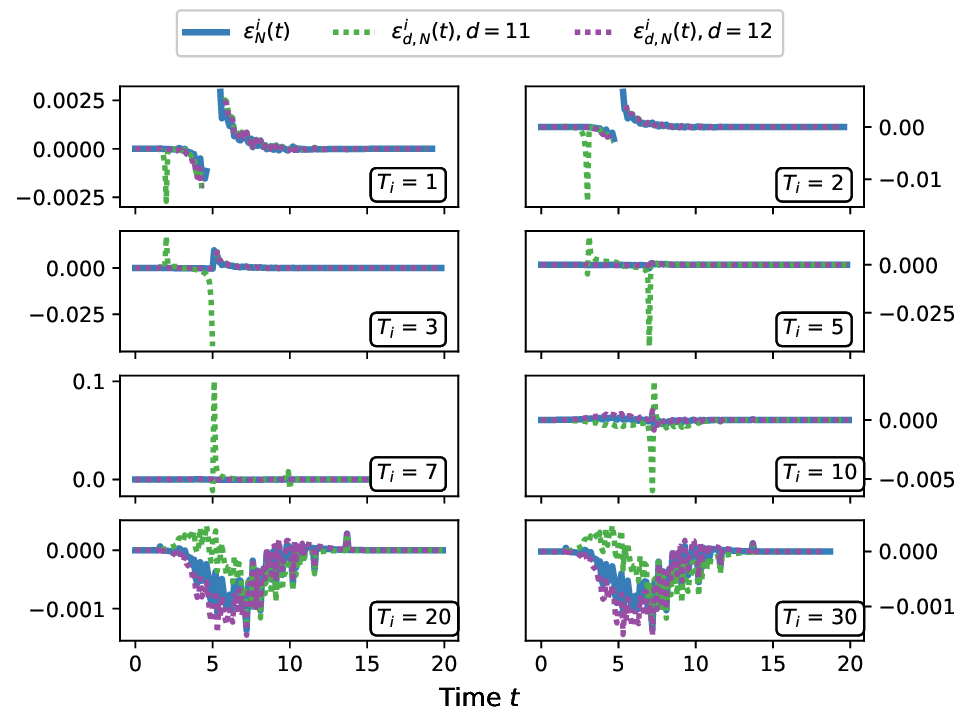}
\caption{Left: The relative errors of the full-order approach $\varepsilon^i_N(t)$ {obtained} with fewer nodes $N=5$ and $N=6$, compared to the relative error of the low-order difference approach, $\varepsilon^i_{d,N}(t)$, for $N=7$, $d=5$.
{Note that the low-order difference approximation outperforms the full-order approximations with reduced values for $N$, even though they have similar numerical complexity.}
\\
Right: The relative approximation errors in a stressed scenario of high volatility $\eta=5\%$ {remain low}. $N=13$, $d\in \{11, 12\}$.}
\label{fig:SwapSensiMethodProof}
\end{figure}

The findings are visualized in the collection of graphs on the left of \Cref{fig:SwapSensiMethodProof}, where the relative errors of these three approaches are compared. From these error plots, we can observe that the low-order difference approximation performs substantially better than both alternative full-order approaches. 
\par
When a choice has to be made on how the number of interpolation nodes is allocated, this implies that it is advantageous to choose a larger value $N$ at the cost of reducing the value $d$, rather than spreading the number of nodes uniformly to $d=N$ with a lower number of nodes $N$. 
{An expected exposure calculation benefits from such a selection, since it utilizes only the unshocked portfolio approximation.}
\par
We conclude the experiments on the single-swap portfolio with a stressed market scenario, where the volatility $\eta$ of the underlying interest rate $r(t)$ is {now $5\%$, increased from previously $2\%$.}
To guarantee that the pricing error of the expected exposure adheres to the threshold, $\varepsilon_\EE < 1$\,bp, the number of interpolation nodes must be increased to $N=13$. 
In the collection of graphs to the right of \Cref{fig:SwapSensiMethodProof}, the corresponding relative errors are displayed for the full-order approach and the low-order difference approaches with $1$ and $2$ points removed, respectively. Magnitude and behaviour of the relative errors appear comparable to the observations made in the previous experiments with an unstressed volatility coefficient, which indicates a robustness of the method.
%
%
%
%
%
\subsection{Large swap portfolio}\label{sec:largeportfolio}
Having examined the properties of the model in a simplified setting, we consider a more realistic portfolio of linear derivatives that involves a larger number of interest rate swaps. The exact composition of this portfolio is given in \ref{sec:numerical-params}. A choice of $N=13$ interpolation nodes ensures an expected exposure estimation error close to the threshold of $1$\, basis point ($\varepsilon_\EE \approx 1.2$\,bp). However, we note that already with a smaller number of nodes $N=9$, the maximal expected exposure error $\varepsilon_\EE$ is smaller than $7$ basis points.
\par
\begin{figure}[]
\centering
\includegraphics[width=.7\textwidth]{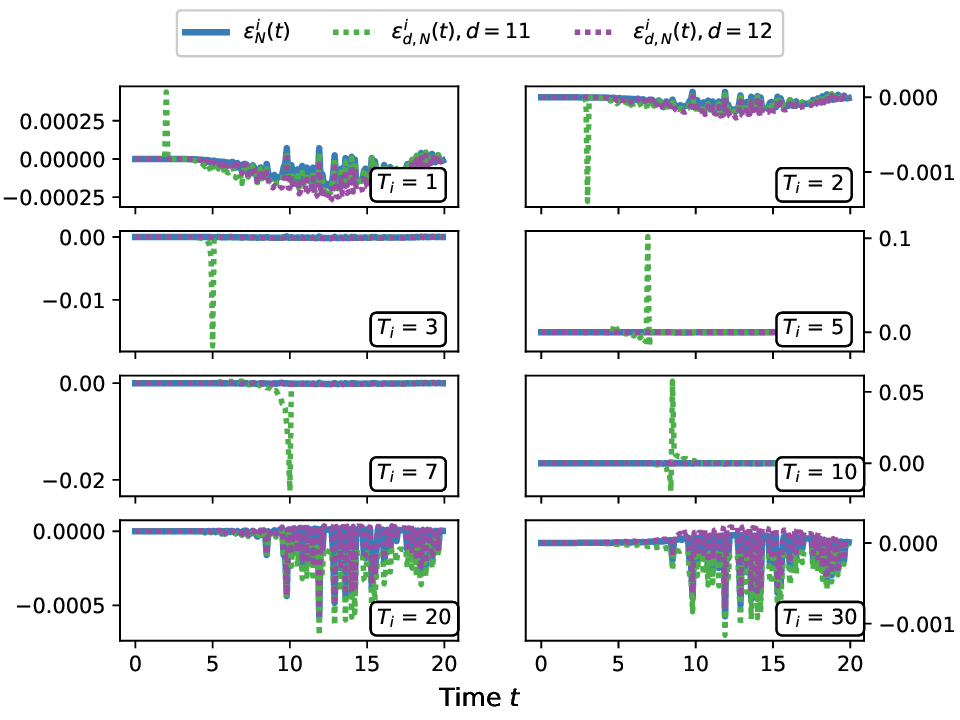}
\caption{Relative approximation errors for the large swap portfolio {remain small, analogous to the single swap portfolio in \Cref{fig:SwapSensi}.}
\\%
$N=13,\ d\in \{11, 12\}$.}
\label{fig:BigSwapSensi}
\end{figure}

In \Cref{fig:BigSwapSensi}, we repeat the previously established experiment to obtain the relative sensitivity approximation errors. As in the previous experiments, we observe extremely small estimation errors. Particularly, it stands out that the largest relative estimation errors continue to be restricted to individual time points.
%
%
%
%
\par
The expected exposures are not a risk metric in themselves, but rather a component of xVA computations. In many of these, the expected exposures are integrated over time, together with weight functions representing risks such as defaults or funding costs, see for example the credit valuation adjustment with wrong-way risk in \eqref{eq:CVA}. 
{We introduce another error measurement that resembles the sensitivities of xVA computations in a general setting. This error, $\zeta^i_d$, relates to the sensitivity of the time-integrated expected exposures,}
\begin{align}
\left| \frac{\partial}{\partial K_i} \int_{t_0}^T\widetilde \EE(t_0, t) \d t - \frac{\partial}{\partial K_i} \int_{t_0}^T\EE(t_0, t) \d t \right| 
& \leq \int_{t_0}^T \left| \frac{\partial}{\partial K_i} \widetilde \EE(t_0, t) - \frac{\partial}{\partial K_i} \EE(t_0, t)\right| \d t \nonumber \\
& \approx \int_{t_0}^T \left|\Psi^i_{d,N}(t) - \Psi^i(t)\right| \d t =: \zeta^i_d.
\end{align}
We recall that $\Psi^i_{N,N}(t) = \Psi^i_N(t)$, so that $\zeta^i_d$ represents the error associated with the full-order approach for $d=N$, and the error associated with the low-order difference approach for $d<N$.
To improve interpretability, we define the normalized versions of these error measurements by
\begin{equation}
\kappa^i_d := \frac{\zeta^i_d}{\int_{t_0}^T \left| \Psi^i(t_0, t) \right| \d t}.
\end{equation}
The findings are arranged in \Cref{tab:integratederror}, where we tabulate the normalized errors $\kappa^i_d$ for $1 \leq i \leq 8$ and a broad range of degrees for the low-order difference approximation, ranging from a linear difference approximation ($d=2$) to the full-order approximation ($d=13$). 
We observe that already with a low-order difference approach in $d=7$ nodes, the normalized error is below $1\%$ for all sensitivities. This translates to $N + d\times n = 69$ exact valuations, %
an improvement of over $40\%$ compared to $117$ valuations in the full-order approach. Both methods improve significantly on the $(n+1)\times M$ evaluations in the classical, exact approach, which range in the tens to hundreds of thousands. {In \Cref{tab:complexity}, an overview over the complexity improvements is provided.}
%
\begin{table}[h]
\footnotesize
\centering
\caption{The normalized, integrated expected exposure sensitivity approximation errors $\kappa^i_d$ for the large swap portfolio. For each shock to $K_i$ (associated with $T_i$), the lowest order $d$ with an approximation error below $1\%$ is highlighted. $N=13$.}
\label{tab:integratederror}

\begingroup
\renewcommand{\arraystretch}{1.2} 
\begin{tabular}{l|llllllll}
\tikz{\node[below left, inner sep=1pt] (def) {$d$};%
      \node[above right,inner sep=1pt] (abc) {$T_i$};%
      \draw (def.north west|-abc.north west) -- (def.south east-|abc.south east);} & 1                & 2                & 3                & 5                & 7                & 10                & 20                & 30                \\ \hline
2      & 1.3E+00          & 2.6E+00          & 2.6E+00          & 1.2E+00          & 1.0E+00          & 5.1E-01          & 1.4E-01          & 2.3E-01          \\
3      & 1.2E-01          & 2.9E-01          & 2.5E-01          & 1.2E-01          & 1.3E-01          & 7.6E-02          & 1.2E-02          & 2.2E-02          \\
4      & 1.0E-01          & 2.3E-01          & 3.6E-01          & 2.0E-01          & 1.8E-01          & 9.4E-02          & 1.3E-02          & 2.8E-02          \\
5      & \textbf{2.0E-03} & \textbf{3.1E-03} & 1.3E-02          & 1.2E-02          & 1.6E-02          & 1.1E-02          & \textbf{1.8E-03} & \textbf{2.8E-03} \\
6      & 1.4E-03          & 1.9E-03          & 1.0E-02          & \textbf{6.6E-03} & 1.2E-02          & 1.1E-02          & 2.2E-03          & 2.9E-03          \\
7      & 4.3E-05          & 1.3E-04          & \textbf{1.3E-03} & 1.4E-03          & \textbf{2.2E-03} & \textbf{1.6E-03} & 3.0E-04          & 4.7E-04          \\
8      & 2.2E-05          & 6.0E-05          & 7.5E-04          & 7.1E-04          & 1.2E-03          & 9.8E-04          & 1.4E-04          & 2.1E-04          \\
9      & 8.0E-06          & 2.3E-05          & 2.6E-04          & 3.9E-04          & 7.9E-04          & 6.5E-04          & 8.8E-05          & 1.3E-04          \\
10     & 1.0E-05          & 1.8E-05          & 2.0E-04          & 2.2E-04          & 6.1E-04          & 6.3E-04          & 1.5E-04          & 1.8E-04          \\
11     & 4.9E-06          & 7.0E-06          & 7.0E-05          & 1.0E-04          & 2.8E-04          & 2.9E-04          & 9.3E-05          & 1.2E-04          \\
12     & 5.7E-06          & 6.6E-06          & 1.1E-05          & 7.9E-06          & 3.3E-05          & 8.4E-05          & 4.5E-05          & 8.6E-05          \\
13     & 3.9E-06          & 4.4E-06          & 7.4E-06          & 4.6E-06          & 2.6E-05          & 4.1E-05          & 4.3E-05          & 7.2E-05         
\end{tabular}
\endgroup
\end{table}
\par
%
\begin{table}[h]
\footnotesize
\centering
\caption{The number of exact valuations at each monitoring date $t$ obtained from the classical approach and the approximation methods. We assume a Monte Carlo simulation with $M=20000$ samples, $N=13$ nodes in the unshocked market and $n=8$ market shocks.}
\label{tab:complexity}
\begingroup
\renewcommand{\arraystretch}{1.2} 
\begin{tabular}{r|cccccc|c|c}
\multicolumn{1}{l|}{}                                                       & \multicolumn{6}{c|}{\textbf{Low-order}}          & \textbf{Full-order} & \textbf{Classical} \\
\multicolumn{1}{l|}{}                                                       & \multicolumn{6}{c|}{$N + dn$}           & $N + Nn$   & $M + Mn$  \\
$d$                                                                         & 7    & 8    & 9    & 10   & 11   & 12   &            &           \\ \hline
\begin{tabular}[c]{@{}r@{}}Number of\\ exact valuations\end{tabular}        & 69   & 77   & 85   & 93   & 101  & 109  & 117        & 180'000   \\
\begin{tabular}[c]{@{}r@{}}Proportion of\\ full-order approach\end{tabular} & 59\% & 66\% & 73\% & 79\% & 86\% & 93\% & 100\%      & $>1500\%$
\end{tabular}
\endgroup
\end{table}
%
%
%
\section{Approximation and sensitivity of Bermudan swaptions}\label{sec:BermSwaptionExperiment}
We now turn our attention to the approximation of exposure and sensitivity of Bermudan swaptions. These are non-linear interest rate derivatives composed of an underlying swap and the optionality of entering this swap at multiple exercise dates. This makes the valuation of a Bermudan swaption an optimal-exercise type problem, where at each exercise date, the (expected) pay-off from exercising the option (i.e., entering the underlying swap) has to be weighed against the continuation of the option to the next exercise date. This results in a high numerical valuation cost, particularly so in exposure simulations. 
\par
Efficient pricing algorithms have received much attention in the literature. In the seminal article of \cite{longstaff2001valuing}, the authors introduce the least squares Monte Carlo (LSMC) algorithm, which can be briefly described as a recursive, backwards search through the option values at the exercise times, where the continuation values {are} approximated {by} linear regressions. This allows for a simulation-based valuation, which only requires one set of paths per valuation time.
Another notable approach has been introduced by \cite{Glau2020}, where an iterative use of Chebyshev interpolation approximates the continuation values and allows for further complexity reduction in the exposure simulation of a Bermudan swaption.
\par
In our outline of a possible approximation for the valuation of a Bermudan swaption $U(t, r(t))$ with underlying interest rate swap $V(t, r(t))$, we assume physical settlement, i.e.\ the Bermudan swaption is replaced by the underlying swap after exercise. Let $S_1, \dots, S_L$, $L\in\mathbb{N}$, be the exercise dates and let $\tau \in \R \cup \{\infty\}$ be the time of exercise, where $\tau =\infty$ indicates no exercise of the optionality. 
Then, the portfolio comprised of this Bermudan swaption is given by
	\begin{equation}
	\Pi(t, r(t)) = 
		\begin{cases}
		U(t, r(t)), & t < \tau, \\
		V(r(t)), & t \geq \tau.
		\end{cases}
	\end{equation}
This combination of valuation functions introduces a new challenge to approximations.
In an exposure simulation path $\omega\in\Omega$, the portfolio value at some point $r(t, \omega)$ may be either the value of the Bermudan swaption, $U(t, r(t, \omega))$ if $t < \tau(\omega)$, or the value of the underlying swap, $V(t, r(t, \omega))$ if the option has been previously exercised, i.e., $t \geq \tau(\omega)$. 
\par
This implies that there can be no single polynomial interpolation of the portfolio value. A possible resolution lies in the construction of two approximation functions $g_U(t, \cdot)$ and $g_V(t, \cdot)$, where $g_U$ approximates the Bermudan swaption value $U$, and $g_V$ the swap value $V$.
\par
Then, the approximation of the portfolio valuation is given by
\begin{equation}\label{eq:gbermportfolio}
g(t, r(t, \omega)) = 
	\begin{cases}
	g_U(t, r(t, \omega)), & t < \tau(\omega), \\
	g_V(r(t, \omega)), & t \geq \tau(\omega),
	\end{cases}
\end{equation}
which implies that the path-wise execution times $\tau(\omega)$ must be tracked in the exposure simulation. A numerically efficient way to estimate these execution times is an initial pass over all exposure simulation paths $r(t, \omega)$, for $t \in [t_0, S_n]$ with the LSMC algorithm, which allows for an approximation of the exercise time $\tau(\omega)$ of each path.
{
\cite{Karlsson2016} remark that this approach constitutes an estimation of an exercise boundary. With such a boundary, it becomes possible to rephrase the exercise decision in terms of a threshold $r^*(t)$ in the underlying short rate, where the option is exercised at time $S_k$ if $r(S_k) < r^*(S_k)$. 
\par
This implies that all inputs to the  optionality component approximation $g_U(t, \cdot)$ come from a truncated distribution $r(t)|\{r(t)<r^*(t)\}$. The corresponding quadrature nodes can be computed by the Golub-Welsch algorithm \citep{golub1969calculation} from the moments of this truncated distribution.}
\par
For our numerical experiment, we consider a Bermudan swaption with exercise dates $S_k = k$ for $k\in\{1, \dots, 5\}$,  strike rate $\bar K$ and as an underlying the swap $V(t)$ described in \Cref{sec:singleswap}. 
Whenever the Bermudan swaption is exercised, it transforms into a swap, {as treated} in the previous section. Therefore, we focus on monitoring dates prior to the expiry of the option, $0 \leq t \leq 5$, and solely on the approximation of the valuation function in unexercised paths, i.e., the approximation $g_U(t, \cdot)$ of the optionality component. For the sensitivity approximations, we restrict the profile to sensitivities with respect to $T_i \in \{1, 2, 3, 5\}$ years, where a large effect on the optionality component can be observed.
\par
\begin{figure}[]
\centering
\includegraphics[width=.49\textwidth]{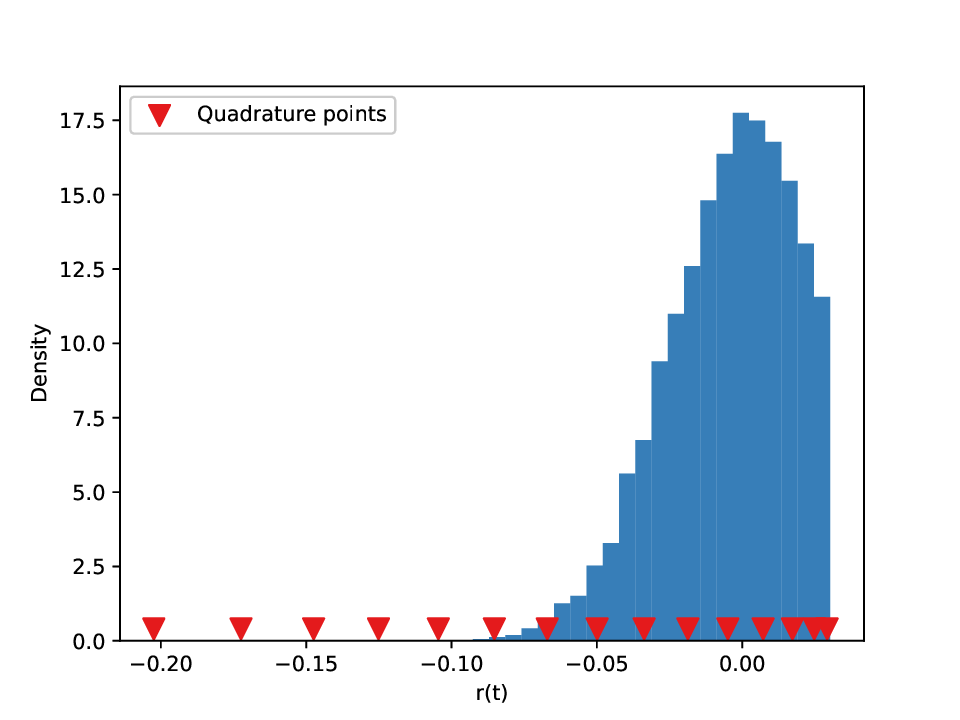}
\includegraphics[width=.49\textwidth]{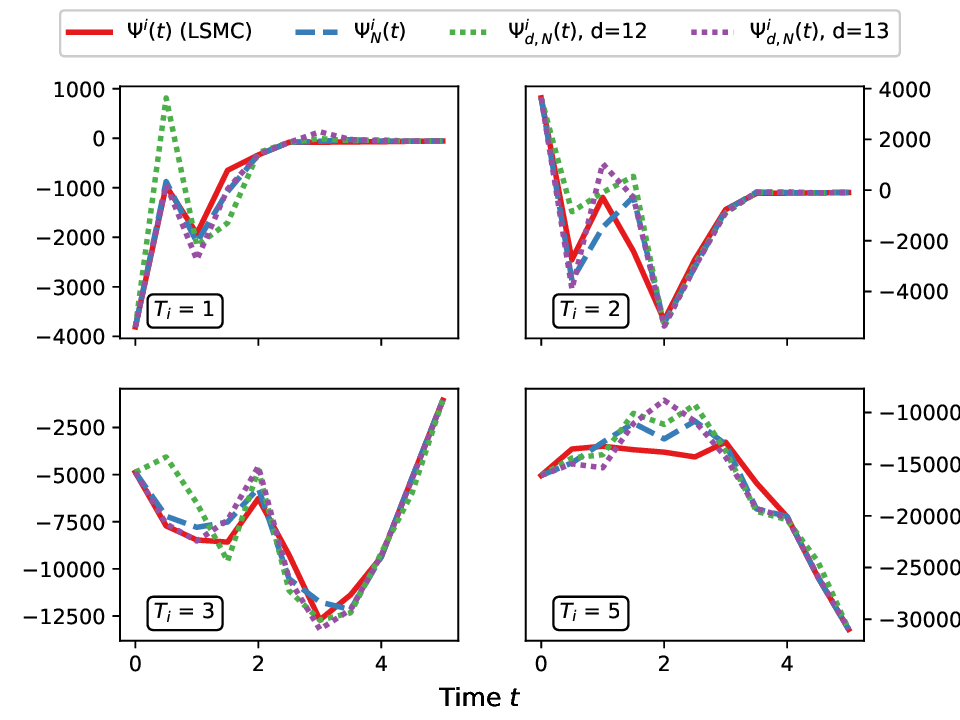}
\caption{Left: Distribution of the risk factor in unexercised paths at time $t=2$ and the corresponding quadrature nodes for $N=15$. Right:  Sensitivity profiles of the Bermudan Swaption, obtained from the full- and low-order approximation methods and the LSMC algorithm.}
\label{fig:Bermudan}
\end{figure}

{As indicated above, the quadrature nodes can be explicitly computed with the moments of the truncated distribution. In this experiment, the interest rate is simulated with the Hull--White dynamics given in \eqref{eq:rt}, so the required moments are those of a truncated normal distribution which can be explicitly computed, see for example \cite{burkardt2014truncated}. 
In the left graph of \Cref{fig:Bermudan}, we display the distribution of the samples $r(t, \omega)$ that are evaluated with the approximation of the optionality component, $g_U(t, \cdot)$, and we show the corresponding choice of quadrature interpolation nodes at time $t=2$.
\par
We find that with these quadrature nodes, a choice of $N=15$ nodes yields a maximal expected exposure error $\varepsilon_\EE < 3$\, bp. 
In the graph on the right of \Cref{fig:Bermudan}, we compare the sensitivity profile obtained from the LSMC algorithm with profile obtained from the full-order and low-order approximation for $N=15$, $d\in\{12, 13\}$. %
In this experiment, the high computational demand of {computing expected exposures with} the LSMC algorithm leads to relatively low accuracy of the `exact' pricing approach, which makes it challenging to draw deep conclusions about the sensitivity approximation error. We recognize the sensitivity profiles computed in \Cref{fig:Bermudan} as plausible and emphasize the large numerical savings provided by the introduced methods.}
%
%
%
\section{Conclusion}\label{sec:conclusion}
{We have shown that the computational demands of expected exposure sensitivity calculations can be greatly reduced in a setting of bump-and-revalue schemes,  used to obtain xVA sensitivities with respect to the yield curve. We showed that the use of quadrature interpolation nodes allows for efficient approximations of the portfolio valuation functions, which is combined with a stochastic collocation technique to obtain the desired sensitivities.}
Two {methods for the determination} of the expected exposure sensitivity have been introduced.
The first, full-order {approach applies equal-degree polynomial approximations to shocked and unshocked valuation functions.}
The second, low-order difference approach offers an even stronger reduction in numerical complexity {by utilizing a reduced order model to appraise the difference introduced to the valuation function by the market shock}.
\par
{These methods can be implemented using basic numerical techniques, the computation of the quadrature nodes requires only matrix operations, and the estimators are obtained using classical polynomial interpolation algorithms.}
\par
The convergence of the low-order difference approximation to the full-order approximation of the sensitivity, and of the full-order approximation of the sensitivity to the classical, exact-valuation-based sensitivity have been {studied}.
\par
Numerical experiments have demonstrated the accuracy {and numerical efficiency} of the method. {Further, experiments} indicate that the low-order difference approach, where an increased degree approximation of the unshocked portfolio is combined with a difference approximation of a lower degree, may be used to improve the accuracy for both the expected exposure simulation and the sensitivity computations.
\par
Applications of approximation techniques in exposure simulations have recently provided fruitful research topics, but further challenges and opportunities remain. The approach to sensitivity computations proposed in this article is based on an interpolation node selection which is entirely domain-driven, that is, the selection of nodes only depends on the distribution of the underlying risk factors. Beyond a certain regularity to ensure the approximability of the portfolio valuation functions, little is assumed about their functional structure. Adaptive algorithms that infer structures such as domains of functional variation from pointwise evaluations may provide a useful improvement in efficiency and complexity, {whilst preserving the advantage of not requiring strong assumptions} about the composition of the portfolios considered.
\section*{Acknowledgments}
{The authors thank an anonymous referee for valuable comments that helped improve this article.}
This research is part of the ABC--EU--XVA project and has received funding from the European Union's Horizon 2020 research and innovation programme under the Marie Sk\l{}odowska--Curie grant agreement No.\ 813261.

\bibliography{Lit}

\begin{thebibliography}{24}
\providecommand{\natexlab}[1]{#1}
\providecommand{\url}[1]{\texttt{#1}}
\expandafter\ifx\csname urlstyle\endcsname\relax
  \providecommand{\doi}[1]{doi: #1}\else
  \providecommand{\doi}{doi: \begingroup \urlstyle{rm}\Url}\fi

\bibitem[Abramowitz and Stegun(1964)]{abramowitz1964handbook}
M.~Abramowitz and I.~A. Stegun.
\newblock \emph{Handbook of mathematical functions with formulas, graphs, and
  mathematical tables}, volume~55.
\newblock US Government printing office, 1964.

\bibitem[Berrut and Trefethen(2004)]{berrut2004barycentric}
J.-P. Berrut and L.~N. Trefethen.
\newblock Barycentric {L}agrange interpolation.
\newblock \emph{SIAM review}, 46\penalty0 (3):\penalty0 501--517, 2004.
\newblock \doi{10.1137/S0036144502417715}.

\bibitem[Brigo and Mercurio(2006)]{brigo2006interest}
D.~Brigo and F.~Mercurio.
\newblock \emph{Interest rate models: theory and practice: with smile,
  inflation, and credit}.
\newblock Springer, Berlin New York, 2006.
\newblock \doi{10.1007/978-3-540-34604-3}.

\bibitem[Burden et~al.(2015)Burden, Faires, and Burden]{burden2015numerical}
R.~L. Burden, J.~D. Faires, and A.~M. Burden.
\newblock \emph{Numerical analysis}.
\newblock Cengage learning, 2015.
\newblock \doi{10.13140/2.1.4830.2406}.

\bibitem[Burkardt(2014)]{burkardt2014truncated}
J.~Burkardt.
\newblock The truncated normal distribution.
\newblock \emph{Department of Scientific Computing Website, Florida State
  University}, 1:\penalty0 35, 2014.

\bibitem[Capriotti et~al.(2017)Capriotti, Jiang, and Macrina]{Capriotti2017}
L.~Capriotti, Y.~Jiang, and A.~Macrina.
\newblock {AAD} and least-square {M}onte {C}arlo: Fast {B}ermudan-style options
  and {XVA} {G}reeks.
\newblock \emph{Algorithmic Finance}, 6\penalty0 (1-2):\penalty0 35--49, Oct.
  2017.
\newblock \doi{10.3233/af-170201}.

\bibitem[Ga{\ss} et~al.(2018)Ga{\ss}, Glau, Mahlstedt, and Mair]{Ga2018}
M.~Ga{\ss}, K.~Glau, M.~Mahlstedt, and M.~Mair.
\newblock Chebyshev interpolation for parametric option pricing.
\newblock \emph{Finance and Stochastics}, 22\penalty0 (3):\penalty0 701--731,
  Apr. 2018.
\newblock \doi{10.1007/s00780-018-0361-y}.

\bibitem[Glau et~al.(2019{\natexlab{a}})Glau, Herold, Madan, and
  P\"{o}tz]{Glau2019b}
K.~Glau, P.~Herold, D.~B. Madan, and C.~P\"{o}tz.
\newblock The {C}hebyshev method for the implied volatility.
\newblock \emph{Journal of Computational Finance}, pages 1--31, Dec.
  2019{\natexlab{a}}.
\newblock \doi{10.21314/jcf.2019.375}.

\bibitem[Glau et~al.(2019{\natexlab{b}})Glau, Mahlstedt, and
  P\"{o}tz]{Glau2019a}
K.~Glau, M.~Mahlstedt, and C.~P\"{o}tz.
\newblock A new approach for {A}merican option pricing: The dynamic {C}hebyshev
  method.
\newblock \emph{{SIAM} Journal on Scientific Computing}, 41\penalty0
  (1):\penalty0 B153--B180, Jan. 2019{\natexlab{b}}.
\newblock \doi{10.1137/18m1193001}.

\bibitem[Glau et~al.(2021)Glau, Pachon, and P\"{o}tz]{Glau2020}
K.~Glau, R.~Pachon, and C.~P\"{o}tz.
\newblock Speed-up credit exposure calculations for pricing and risk
  management.
\newblock \emph{Quantitative Finance}, 21\penalty0 (3):\penalty0 481--499,
  2021.
\newblock \doi{10.1080/14697688.2020.1781236}.

\bibitem[Gnoatto et~al.(2020)Gnoatto, Picarelli, and Reisinger]{Gnoatto2020}
A.~Gnoatto, A.~Picarelli, and C.~Reisinger.
\newblock Deep {xVA} solver -- a neural network based counterparty credit risk
  management framework, 2020.
\newblock URL \url{https://arxiv.org/abs/2005.02633}.

\bibitem[Golub and Welsch(1969)]{golub1969calculation}
G.~H. Golub and J.~H. Welsch.
\newblock Calculation of {G}auss quadrature rules.
\newblock \emph{Mathematics of computation}, 23\penalty0 (106):\penalty0
  221--230, 1969.

\bibitem[Green(2015)]{green2015xva}
A.~Green.
\newblock \emph{XVA: Credit, Funding and Capital Valuation Adjustments}.
\newblock John Wiley \& Sons, 2015.
\newblock \doi{10.1002/9781119161233}.

\bibitem[Gregory(2020)]{gregory2020xva}
J.~Gregory.
\newblock \emph{The xVA Challenge: Counterparty Risk, Funding, Collateral,
  Capital and Initial Margin}.
\newblock John Wiley \& Sons, 2020.
\newblock \doi{10.1002/9781119508991}.

\bibitem[Grzelak(2022)]{grzelak2022sparse}
L.~A. Grzelak.
\newblock Sparse grid method for highly efficient computation of exposures for
  {xVA}.
\newblock \emph{Applied Mathematics and Computation}, 434:\penalty0 127446,
  2022.
\newblock \doi{10.1016/j.amc.2022.127446}.

\bibitem[Grzelak et~al.(2018)Grzelak, Witteveen, Su\'arez-Taboada, and
  Oosterlee]{GrzelakColl}
L.~A. Grzelak, J.~Witteveen, M.~Su\'arez-Taboada, and C.~Oosterlee.
\newblock The stochastic collocation {M}onte {C}arlo sampler: highly efficient
  sampling from expensive distributions.
\newblock \emph{Quantitative Finance}, 19:\penalty0 1--18, 06 2018.
\newblock \doi{10.1080/14697688.2018.1459807}.

\bibitem[Hagan and West(2006)]{hagan2006interpolation}
P.~S. Hagan and G.~West.
\newblock Interpolation methods for curve construction.
\newblock \emph{Applied Mathematical Finance}, 13\penalty0 (2):\penalty0
  89--129, 2006.
\newblock \doi{10.1080/13504860500396032}.

\bibitem[Huge and Savine(2017)]{Huge2017}
B.~N. Huge and A.~Savine.
\newblock {LSM} reloaded - differentiate {xVA} on your {iPad} mini.
\newblock \emph{{SSRN} Electronic Journal}, 2017.
\newblock \doi{10.2139/ssrn.2966155}.

\bibitem[Hull and White(1990)]{hull1990pricing}
J.~Hull and A.~White.
\newblock Pricing interest-rate-derivative securities.
\newblock \emph{The Review of Financial Studies}, 3\penalty0 (4):\penalty0
  573--592, 1990.
\newblock \doi{10.1093/rfs/3.4.573}.

\bibitem[Karlsson et~al.(2016)Karlsson, Jain, and Oosterlee]{Karlsson2016}
P.~Karlsson, S.~Jain, and C.~W. Oosterlee.
\newblock Fast and accurate exercise policies for {B}ermudan swaptions in the
  {LIBOR} market model.
\newblock \emph{International Journal of Financial Engineering}, 03\penalty0
  (01):\penalty0 1650005, Mar. 2016.
\newblock \doi{10.1142/s2424786316500055}.

\bibitem[Longstaff and Schwartz(2001)]{longstaff2001valuing}
F.~A. Longstaff and E.~S. Schwartz.
\newblock Valuing {A}merican options by simulation: a simple least-squares
  approach.
\newblock \emph{The Review of Financial Studies}, 14\penalty0 (1):\penalty0
  113--147, 2001.
\newblock \doi{10.1093/rfs/14.1.113}.

\bibitem[Oosterlee and Grzelak(2019)]{GrzelakOosterlee}
C.~W. Oosterlee and L.~A. Grzelak.
\newblock \emph{Mathematical Modeling and Computation in Finance: With
  Exercises and Python and Matlab Computer Codes}.
\newblock World Scientific Publishing Company, 2019.
\newblock \doi{10.1142/q0236}.

\bibitem[Trefethen(2019)]{trefethen2019approximation}
L.~N. Trefethen.
\newblock \emph{Approximation Theory and Approximation Practice, Extended
  Edition}.
\newblock SIAM, 2019.
\newblock \doi{10.1137/1.9781611975949}.

\bibitem[Xiu and Hesthaven(2005)]{XiuHighOrderCollocation}
D.~Xiu and J.~S. Hesthaven.
\newblock High-order collocation methods for differential equations with random
  inputs.
\newblock \emph{SIAM Journal on Scientific Computing}, 27\penalty0
  (3):\penalty0 1118--1139, 2005.
\newblock \doi{10.1137/040615201}.

\end{thebibliography}

%
\appendix
\section{{Extension to more general xVA models on the example of wrong-way risk}}\label{sec:WWR}
In the following, we demonstrate how portfolio valuation approximation can be applied to more complex xVA models, by example of the unilateral CVA with wrong-way risk (WWR).
\par
Generally, the CVA is constructed from three components. The \emph{probability of default} of the counterparty (PD), the \emph{expected (positive) exposure} (EE) and the \emph{loss given default} (LGD), which describes what percentage of the exposure is lost in a default event. A common and simplifying CVA approximation {is based on the assumption of independence between exposure and the probability of default} (see, for example, \cite{gregory2020xva}). %
The resulting credit valuation adjustment is given by
\begin{equation}
\CVA_1(t_0) := \mathrm{LGD}\cdot \sum_{k=1}^R \EE(t_0, t_k) \mathrm{PD}(t_{k-1}, t_k),
\end{equation}
{where $t_0 <  \dots < t_R$, $R\in\mathbb{N}$, is a discretization of the time horizon.}
The independence assumption which allows for factorization of EE and PD is not always warranted, as the exposure may rely on the same risk factors which influence the default risk. Depending on the sign of correlation, this is known as wrong-way risk (when exposure increases together with the probability of default) or right-way risk (when exposure decreases as the probability of default rises). Let $\tau$ denote the time of counterparty default and assume that at the time of valuation $t_0$, the default has not yet occurred, $\tau > t_0$. 
Then, a more general formulation 
 of the CVA is given by 
\begin{equation}
\CVA_2(t_0) := \mathrm{LGD}\cdot  \E_{t_0}^\Q\Bigl[\frac{B(t_0)}{B(\tau)} \1_{\{\tau \leq T\}} V^+(\tau) \Bigr].
\end{equation}
{One way to introduce correlations is through} stochastic hazard rates $y(t)$, which are continuous stochastic processes that describe the {intensity default process}. Correlating these hazard rates with the underlying is one way to induce an analytically solvable dependency structure between default times and exposure. Analogous to the stochastic intensity model {in Chapter 22.7} of \cite{brigo2006interest}, it is possible to show that the previous equation can be rewritten as
\begin{equation}
\CVA_2(t_0)  = \mathrm{LGD}\cdot \int\limits_{t_0}^T \E_{t_0}^\Q\Bigl[\frac{B(t_0)}{B(s)} \Bigl(\e^{-\int_{t_0}^s y_u \d u} y_s\Bigr) V^+(s) \Bigr] \d s. \label{eq:CVA}
\end{equation}
This expected value contains an additional {factor} corresponding to the probability of default, $D(s) := \exp(-\int_{t_0}^s y_u \d u) y_s$, which is governed by the stochastic hazard rates. The asset valuation, however, is not impeded. Therefore, a Monte Carlo experiment analogous to \eqref{eq:ee} can be conducted, {where an additional risk factor $y(t)$ is simulated}. The {approximation techniques discussed in this article} can be applied in the usual way to replace the exact portfolio valuations $V$ with a polynomial approximation function $g$, to the effect of the same efficiency improvements as before.
%
%
%
%
\section{Explicit discount factor sensitivity}\label{sec:appx1}
We consider the derivative of the discount factor
\begin{equation}
\frac{\partial}{\partial K_i}  \frac{B(t_0; \omega)}{B(t; \omega)}  = \frac{\partial}{\partial K_i} \exp\left(-\int_{t_0}^t r(s; \omega)\d s\right).
\end{equation}
In the following, we will consider the explicit choice of the 1-factor Hull--White model for the interest rate $r(t)$, %
but the methodology can be extended to many other exogenous interest rate models (see \cite{brigo2006interest} for an overview of such models).
Then, the short rate $r(t)$ can be explicitly expressed in terms of a stochastic process $u(t)$, and a purely deterministic function $\psi(t)$, such that at any time it holds $r(t; \omega) = u(t; \omega) + \psi(t)$.
 The deterministic function $\psi(t)$ imposes the term structure of interest rates, i.e.\ the {initial} yield curve, onto the model, whereas the stochastic process $u(t)$ only depends on the yield curve through its parameter calibration procedure. As explained in Chapter 21 of \cite{green2015xva}, the remaining model parameters are not usually recalibrated when sensitivities to the yield curve are computed. Under this assumption, it holds that
\begin{align}\label{eq:derivbankaccountdecomp}
\frac{\partial \exp(-\int_{t_0}^t r(s; \omega)\d s)}{\partial K_i} 
&= \exp\left(\int_{t_0}^t -u(s; \omega)\d s\right) \frac{\partial \exp(\int_{t_0}^t - \psi(s)\d s)}{\partial K_i}	\nonumber \\
&= \exp\left(\int_{t_0}^t -u(s; \omega)\d s\right) \exp\left( \int_{t_0}^t - \psi(s)\d s \right) \left(-\int_{t_0}^t \frac{\partial \psi(s)}{\partial K_i}  \d s \right).
\end{align}
The deterministic function $\psi(t)$ is explicitly known, in this example of the 1-factor Hull--White model it is given for $t_0 = 0$ by
\begin{equation}
\psi(t) =  f^r(0, t) + \exp(-\lambda t)  \frac{\eta^2}{\lambda^2} (\mathrm{cosh}(\lambda t) - 1 ),
\end{equation}
where $\lambda$ is the speed of mean reversion parameter, $\eta$ is the volatility parameter, and $f^r(0, t)$ %
is the instantaneous forward rate associated with the model. Under the yield curve, respectively its associated zero-coupon bond curve $P(0, \cdot)$ given in \eqref{eq:discountcurve}, the instantaneous forward rate at time $t$ {equals}
\begin{equation}
f^r(0, t) = \frac{-\partial \log(P(0, t))}{\partial t}.
\end{equation}
Since the model parameters $\lambda, \eta$ are not recalibrated to the change in the yield curve, it can be shown that it holds
\begin{equation}\label{eq:detfuncderivative}
\frac{\partial \psi(t)}{\partial K_i} = \frac{\partial f^r(t)}{\partial K_i} =  \frac{\partial \left( -\frac{\partial \log(P(0, t))}{\partial t}    \right)}{\partial K_i} = -\frac{\partial}{\partial t} \left( \frac{1}{P(0, t)} \frac{\partial P(0, t)}{\partial K_i} \right).
\end{equation}
The derivative with respect to the market quote, $\partial P(0, t) / \partial K_i$ is typically not analytically available since the zero-coupon bond curve $P(0, t)$ is obtained from the market quotes by means of a curve-fitting scheme. 
%
%
{
\section{Sensitivity in the yield curve construction}\label{appx:yieldcurve}
As noted below \eqref{eq:detfuncderivative}, the derivatives of the zero-coupon bonds with respect to the market quote $K_i$ are not generally available, since these bonds are often obtained through contrived curve-fitting schemes. 
In generality, each zero-coupon bond $P(0, t)$ depends on all of the market instruments $\varphi_1, \dots, \varphi_n$ (given in \Cref{dfn:constructinginstruments}) on which the yield curve is based.
 We may emphasize this by writing
\begin{equation}
P_t(\varphi_1, \dots, \varphi_n) := P(0, t).
\end{equation}
This implies that the inner derivative in \eqref{eq:detfuncderivative} is expressed by
\begin{equation}
\frac{\partial P(0, t)}{\partial K_i} =  \frac{\partial P_t(\varphi_1, \dots, \varphi_n)}{\partial K_i} = \sum\limits_{j=1}^n \frac{\partial P_t(\varphi_1, \dots, \varphi_n)}{\partial \varphi_j} \frac{\partial \varphi_j}{\partial K_i}.
\end{equation}
Importantly, the effect of each market instrument on each zero-coupon bond is model-dependent. For example, when the yield curve is constructed with the Newton-Raphson method, a Jacobian of the form 
\begin{equation}
J_\mathbf{\varphi} = \begin{pmatrix}
\frac{\partial \varphi_1}{\partial P(0, S_1)} & \cdots & \frac{\partial  \varphi_1}{\partial P(0, S_n)}\\
\vdots & \ddots & \vdots \\
\frac{\partial  \varphi_n}{\partial P(0, S_1)} & \cdots & \frac{\partial  \varphi_n}{\partial P(0, S_n)}
\end{pmatrix}
\end{equation}
is obtained, where each maturity $S_j$ is a spine point associated with exactly one market instrument $\varphi_j$, $j\in\{1, \dots, n\}$. Under some restrictions on the choice of market instruments such that the inverse function theorem holds for $J_\mathbf{\varphi}$, the inverse Jacobian 
\begin{equation}
J_{P} = 
\begin{pmatrix}
\frac{\partial P(0, S_1)}{\partial \varphi_1} & \cdots & \frac{\partial P(0, S_1)}{\partial \varphi_n}\\
\vdots & \ddots & \vdots \\
\frac{\partial P(0, S_n)}{\partial \varphi_1} & \cdots & \frac{\partial P(0, S_n)}{\partial \varphi_n}
\end{pmatrix}
\end{equation}
can be found explicitly with the relation $J_{P} = J_{\varphi}^{-1}$.
It can not be guaranteed that every maturity $t$, for which a zero-coupon bond $P(0, t)$ is required, is also an element of the spine points $\{S_1, \dots, S_n\}$. This brings about the need for another approximation.
Direct interpolation between the interpolation points
$
\bigl\{\frac{\partial P(0, S_\ell)}{\partial \varphi_j}\colon \ell \in \{1, \dots, n\} \bigr\}
$
often does not admit good results. In these cases, it is preferable to deploy another chain rule, 
\begin{equation}\label{eq:discountchainrule}
\frac{\partial P(0, t)}{\partial \varphi_j} = \sum_{j=1}^n \frac{\partial P(0, t)}{\partial P(0, S_\ell)} \frac{\partial P(0, S_\ell)}{\partial \varphi_j},
\end{equation}
and interpolate between the missing terms $\partial P(0, t)/\partial P(0, S_\ell)$ instead. Clearly, even in such a semi-analytical approach, the yield curve construction methodology effects the sensitivities. The same holds in direct ``bump-and-revalue'' schemes, as used within this article, directly through the construction of the shocked and unshocked yield curves.
}
%
\section{Parameters of the numerical experiments}\label{sec:numerical-params}
In the numerical experiments, a yield curve is constructed based on (artificial) market quotes for interest rate swaps, given in \Cref{tab:constructing-instruments}. The composition of the large swap portfolio is specified in \Cref{tab:swapportfolio}.

\begin{table}[h]
\footnotesize
\centering
\caption{Market instruments (interest rate swaps) used in the yield curve construction.}
\label{tab:constructing-instruments}
\begin{tabular}{rllllllll}
$i:$ & 1 & $2$    & $3$    & 4    & 5    & 6   & 7   & 8   \\
Maturity $T_i:$                                                   & 1 & $2$    & $3$    & 5    & 7    & 10   & 20   & 30   \\
\begin{tabular}[c]{@{}l@{}}Swap rate $K_i$ (in \%):\end{tabular} & 0.04       & 0.16 & 0.31 & 0.81 & 1.28 & 1.62 & 2.22 & 2.30
\end{tabular}
\end{table}
\begin{table}[h]
\centering
\footnotesize
\caption{Specification of the interest rate swaps comprising the large swap portfolio.}
\label{tab:swapportfolio}
\begin{tabular}{rrrrrr}%
Sign & Notional & Fixed Rate & Maturity & Start Date & Payments per Year \\
-1   & 10000    & 0.022      & 20       & 0          & 2.0               \\
-1   & 8333     & 0.042      & 20       & 0          & 2.0               \\
-1   & 8333     & 0.042      & 21       & 0          & 1.9               \\
1    & 8333     & 0.042      & 24       & 0          & 1.7               \\
1    & 8333     & 0.042      & 17       & 0          & 2.4               \\
1    & 8333     & 0.042      & 26       & 0          & 1.5               \\
1    & 8333     & 0.042      & 19       & 5          & 2.9               \\
1    & 8333     & 0.042      & 40       & 10         & 1.3               \\
-1   & 8333     & 0.042      & 19       & 3          & 2.5               \\
-1   & 8333     & 0.042      & 20       & 7          & 3.1               \\
1    & 8333     & 0.042      & 20       & 2          & 2.2               \\
-1   & 8333     & 0.042      & 20       & 0          & 2.0              \\
-1   & 8333     & 0.042      & 20       & 0          & 2.0              \\
\end{tabular}
\end{table}

\end{document}